\documentclass[aps,prx,10pt,twocolumn,amsmath,amssymb,longbibliography,superscriptaddress,floatfix]{revtex4-2}

\usepackage{graphicx}
\usepackage{bm}
\usepackage{booktabs}
\usepackage{placeins}
\usepackage{needspace}
\usepackage{xcolor}
\usepackage[normalem]{ulem}
\usepackage{environ}
\definecolor{georgegreen}{RGB}{0,110,55}
\NewEnviron{strikethrough}{\expandafter\sout\expandafter{\BODY}}[]

\usepackage[colorlinks=true,allcolors=blue]{hyperref}
\usepackage{orcidlink}
\usepackage{times,txfonts}
\newcommand{\ZZ}{\mathbb{Z}_2}

\newcommand{\ii}{\mathrm{i}}
\newcommand{\xiR}{\xi_R}

\begin{document}
\title{Ground-State Spin Squeezing and Collective Excitations in $\mathbb{Z}_2$-Symmetric Systems}

\author{Arkadiusz Kosior\,\orcidlink{0000-0002-5039-1789}}
\affiliation{Institute of Physics, Maria Curie-Sk\l odowska University, 20-031 Lublin, Poland}
\affiliation{Institut f\"ur Theoretische Physik, Universit\"at Innsbruck, Technikerstra{\ss}e 21a, A-6020 Innsbruck, Austria}

\author{George Mihailescu\,\orcidlink{0000-0002-0048-9622}}
\affiliation{School of Physics, University College Dublin, Belfield, Dublin 4, Ireland}
\affiliation{Centre for Quantum Engineering, Science, and Technology, University College Dublin, Dublin 4, Ireland}

\author{Karol Gietka\,\orcidlink{0000-0001-7700-3208}}
\email{karol.gietka@uibk.ac.at}
\affiliation{Institut f\"ur Theoretische Physik, Universit\"at Innsbruck, Technikerstra{\ss}e 21a, A-6020 Innsbruck, Austria}

\date{\today}

\begin{abstract}
{Ground-state spin squeezing is a resource for quantum-enhanced measurements, but is commonly associated with a closing excitation gap near a quantum critical point. A closing gap can complicate state preparation and reduce thermal robustness. Here we identify the physical mechanism linking squeezing to collective excitations and establish when stronger squeezing accompanies a larger many-body gap. Using the anisotropic Lipkin--Meshkov--Glick model, we show that, for a quadratic collective mode, the ratio of two transverse fluctuation costs determines squeezing, whereas their geometric mean sets the collective gap. The pair terms responsible for unequal costs are allowed, but not guaranteed, by $\mathbb Z_2$ parity. To extend this picture beyond a single mode, we derive an exact relation between squeezing, polarization, and excitation energies weighted by their collective-spin transition strengths. For XXZ chains, weak-coupling and spin-wave analyses explain how squeezing can develop while both the collective and many-body gaps increase. Exact finite-size calculations for power-law interactions, including the nearest-neighbor limit, confirm a negative-coupling branch on which the two gaps coincide and increase as squeezing improves. For nearest-neighbor interactions, this branch also retains squeezing at higher temperatures than a gap-closing branch with comparable ground-state squeezing. In the Ising chain, by contrast, a staggered excitation can lower the many-body gap without directly contributing to collective fluctuations. Our results explain why ground-state squeezing can accompany either a closing or an opening collective gap, and identify a gap-opening branch with enhanced thermal robustness and the potential for faster adiabatic preparation.}
\end{abstract}

\maketitle

\section{Introduction}
\label{sec:introduction}

Spin squeezing is a key resource for quantum-enhanced metrology, enabling measurement sensitivities beyond the standard quantum limit while also providing a practical witness of many-particle entanglement~\cite{KitagawaUeda1993,Wineland1994,Sorensen2001,Gross2012,MaReview2011,TothApellaniz2014,Pezze2018,Giovannetti2011}. Experimentally, it has been generated using atomic collisions, cavity-mediated interactions, measurement backaction, Rydberg dressing, and dipolar interactions~\cite{Esteve2008,Riedel2010,Muessel2014,Bohnet2014,Hosten2016,Eckner2023,Lee2025,Douglas2025}. One route to squeezing is to prepare an interacting many-body ground state whose correlations could already reduce collective spin fluctuations. This approach is particularly appealing near a quantum phase transition, where ground-state correlations are often enhanced~{\cite{OsborneNielsen2002,Osterloh2002,VidalLatorre2003,AmicoReview2008,MontenegroReview2025}}. In several models, stronger squeezing accompanies the closing of a collective excitation gap~\cite{MaWang2009,FrerotRoscilde2018,Garbe2020,Mirkhalaf2020,Mirkhalaf2021,ShaoFu2024,Zhang2024}. This association, however, does not explain why squeezing develops or whether gap closing is necessary. The distinction also matters for preparing and maintaining the squeezed state. When the lowest excitation gap closes, adiabatic preparation can become slower and thermal excitations easier to populate~\cite{Rams2018,GietkaSpeeding2022,GietkaRuksBusch2022,Hotter2024,OstermannGietka2024,MihailescuCampbellGietka2025}. It is therefore natural to ask whether this trade-off is unavoidable. \emph{What connects ground-state spin squeezing to the excitation spectrum, and can squeezing increase while the relevant excitation becomes more energetic?}

Gap-opening squeezing has already been proposed as a route to faster state preparation in collective models~\cite{GietkaSpeeding2022}, and anti-critical quantum metrology has extended this perspective to sensing with an increasing excitation gap~\cite{MihailescuGietkaAntiCritical2026}. The role of pair-flip processes in producing ground-state squeezing is also established perturbatively~\cite{Roscilde2021}. The question here is how this picture extends beyond a single collective mode.

We begin with the anisotropic Lipkin--Meshkov--Glick (LMG) model, where squeezing and the collective excitation energy can be related analytically~\cite{LMG1965,DusuelVidal2004}. Varying its anisotropy makes the role of pair processes and their symmetry constraints explicit. This requires distinguishing the symmetry of the Hamiltonian from that of the selected ground state, particularly when interpreting squeezing in systems with continuous spin-rotation symmetry~\cite{Comparin2021,Comparin2022,Caleca2025,CalecaLuttinger2025}. {We then use an exact relation involving all excitations coupled to the total spin to compare XXZ and Ising chains. This separates the spectral changes associated with squeezing from those that raise the lowest excitation energy and improve thermal robustness.}

\begin{figure*}[htb!]
\centering
\includegraphics[width=\textwidth]{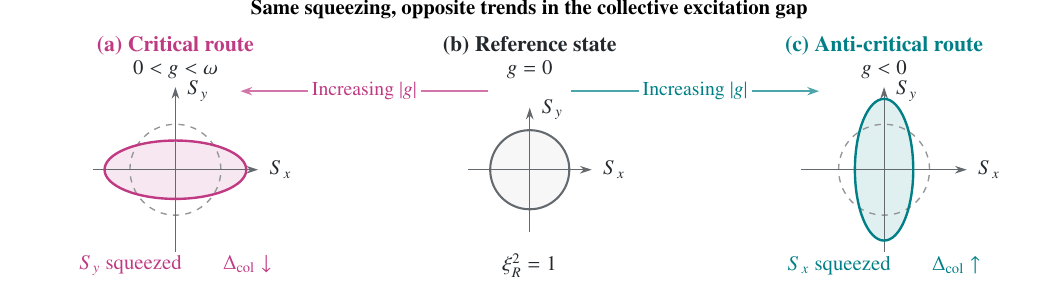}
\caption{Two routes to ground-state squeezing in a $\mathbb Z_2$-symmetric collective mode. Panels (a)--(c) schematically show transverse fluctuations about a mean spin directed along the negative-$z$ axis. The squeezed ellipses preserve parity under $(S_x,S_y)\to(-S_x,-S_y)$. This symmetry permits unequal transverse variances but does not by itself guarantee squeezing. Dashed circles indicate the coherent-spin reference, with $\xi_R^2=1$. As $|g|$ increases from zero, positive coupling squeezes $\hat S_y$ while reducing the collective gap, whereas negative coupling squeezes $\hat S_x$ while increasing it. The ellipses illustrate equal squeezing despite these opposite gap trends.}
\label{fig:mechanism_v16}
\end{figure*}

We focus on equilibrium ground-state squeezing and its thermal robustness, rather than dynamical generation through sudden quenches or open-system protocols~\cite{MunozArias2023,Bilitewski2021,Franke2023,Bornet2023,Block2024,Debecker2025,Jiao2026,Bahamnia2026,poggi,Plodzien2026}. Our aim is to understand which fluctuations generate squeezing and determine when the same interactions also raise the lowest excitation energy and improve thermal robustness. Figure~\ref{fig:mechanism_v16} summarizes the two squeezing routes and their opposite trends in the collective excitation gap.

\section{Collective mechanism in the anisotropic LMG model}
\label{sec:lmg}

\begingroup\widowpenalty=10000
We begin by relating the energy costs of transverse spin fluctuations to ground-state squeezing and the collective excitation gap. The LMG model provides a simple setting because small fluctuations of its total spin reduce to a single mode. Once this connection is established, the familiar gap-closing regime and the gap-opening regime follow from changing the same interaction in opposite directions.
\par\endgroup

\subsection{LMG Hamiltonian, symmetries, and phase structure}

The all-to-all interactions of the LMG model let us write its Hamiltonian entirely in terms of the total spin (we set $\hbar=1$),
\begin{equation}
\hat H_{\mathrm{LMG}}
=2\omega \hat S_z
-\frac{2g}{N}\left(\hat S_x^2+\gamma \hat S_y^2\right),
\label{eq:lmg}
\end{equation}
where $N$ is the number of spin-$1/2$ particles and $\hat S_\mu=\tfrac12\sum_{i=1}^{N}\hat\sigma_i^\mu$ are the collective spin operators, with $\hat\sigma_i^\mu$ the Pauli operators and $\mu=x,y,z$. The positive parameter $\omega$ sets the field energy scale along $z$, so an isolated spin has a level splitting $2\omega$. The interaction strength $g$ can be positive or negative. The dimensionless anisotropy $\gamma$ sets the relative strength of the $y$ interaction. We take $0\leq\gamma\leq1$, from Ising interactions at $\gamma=0$ to equal $x$ and $y$ interactions at $\gamma=1$. The factor $1/N$ makes the interaction energy grow proportionally to $N$. Since the Hamiltonian commutes with $\hat{\boldsymbol S}^{\,2}$, states with different total-spin quantum numbers $S$ can be treated separately. We work in the fully symmetric sector with maximal total spin $S=N/2$.

Before expanding around a mean-spin direction, it is useful to identify which processes the interaction permits. For every $\gamma$, the Hamiltonian commutes with the spin-parity operator
\begin{equation}
\hat\Pi_z
=e^{\ii\pi(\hat S_z+N/2)}
=\prod_{i=1}^{N}(-\hat\sigma_i^z),
\end{equation}
which changes the signs of $\hat S_x$ and $\hat S_y$ while leaving $\hat S_z$ unchanged. For nonzero $g$ and $\gamma\neq1$, the Hamiltonian preserves this $\mathbb Z_2$ symmetry but does not conserve $\hat S_z$. At $\gamma=1$, it also conserves magnetization, $[\hat H_{\mathrm{LMG}},\hat S_z]=0$, and is invariant under continuous $U(1)$ rotations about $z$.

To see what changes with the anisotropy, we separate the interaction into
\begin{equation}
\hat S_x^2+\gamma\hat S_y^2
=\frac{1+\gamma}{2}\left(\hat S_x^2+\hat S_y^2\right)
+\frac{1-\gamma}{2}\left(\hat S_x^2-\hat S_y^2\right).
\label{eq:channels}
\end{equation}
The first term conserves $\hat S_z$, whereas the second,
\begin{equation}
\hat S_x^2-\hat S_y^2
=\frac12\left(\hat S_+^2+\hat S_-^2\right),
\qquad
\hat S_\pm=\hat S_x\pm\ii\hat S_y,
\end{equation}
raises or lowers its eigenvalue by two units. Relative to a state polarized along negative $z$, these terms create or remove spin-flip excitations in pairs. Such processes preserve parity. Their strength is proportional to $g(1-\gamma)$ and vanishes at the isotropic point $\gamma=1$.

We next identify the mean-spin direction around which these fluctuations develop. For $0\leq\gamma<1$ and $0<g<\omega$, the mean-field ground state is polarized along negative $z$, with $\langle\hat{\boldsymbol S}\rangle_{\mathrm{MF}}=(0,0,-N/2)$. Here the subscript $\mathrm{MF}$ denotes the mean-field value. This is the normal phase, in which the ground state preserves parity. Increasing $g$ destabilizes this direction. Beyond $g=\omega$, or $0<\omega/g<1$, the mean-field energy has two ferromagnetic minima,
\begin{equation}
    \langle\hat{\boldsymbol S}\rangle_{\mathrm{MF}}
    =\frac{N}{2}
    \left(
    \pm\sqrt{1-(\omega/g)^2},
    0,
    -\omega/g
    \right),
    \end{equation}
related by parity. In the thermodynamic limit, the transition at $\omega/g=1$ is continuous and either minimum can be selected through spontaneous $\mathbb Z_2$ symmetry breaking.

Negative coupling instead raises the mean-field energy associated with transverse spin components, so the mean spin remains along negative $z$. The two polarized branches, $0<g<\omega$ and $g<0$, therefore have the same mean-spin direction. Their different squeezing must come from fluctuations around that direction, which we can describe using the same expansion. In particular, $\omega/g\to0^-$ means $g\to-\infty$ at fixed $\omega>0$, not another transition at a finite coupling. In the ordered phase, the expansion must instead follow one of the selected minima (see Appendix~\ref{app:ferromagnetic} for details).

\subsection{Quadratic fluctuation Hamiltonian and collective excitation gap}

The pair interaction mixes the polarized state with states containing flipped spins. When the number of flipped spins is small compared with $N$, we can describe these fluctuations by a bosonic annihilation operator $\hat a$, with $[\hat a,\hat a^\dagger]=1$. The Holstein--Primakoff representation~\cite{HolsteinPrimakoff1940} gives
\begin{align}\label{eq:hp}
\hat S_z&=-\frac N2+\hat a^\dagger\hat a,\nonumber\\
\hat S_x&=\frac{\sqrt N}{2}(\hat a+\hat a^\dagger)+O(N^{-1/2}),\nonumber\\
\hat S_y&=\frac{\sqrt N}{2\ii}(\hat a^\dagger-\hat a)+O(N^{-1/2}).
\end{align}
Each boson raises $\hat S_z$ by one unit, reducing the magnitude of the negative-$z$ polarization. The expansion requires $\langle\hat a^\dagger\hat a\rangle\ll N$. Keeping the quadratic terms in Eq.~\eqref{eq:lmg} gives {the large-$N$ fluctuation Hamiltonian~\cite{DusuelVidal2005}},
\begin{equation}
\hat H_{\mathrm{LMG}}=E_{\mathrm{ref}}+A\hat a^\dagger\hat a
+\frac B2(\hat a^2+\hat a^{\dagger2})+O(N^{-1}),
\label{eq:quadratic}
\end{equation}
with
\begin{equation}
A=2\omega-g(1+\gamma),\qquad B=-g(1-\gamma),
\label{eq:AB}
\end{equation}
and $E_{\mathrm{ref}}=-N\omega-g(1+\gamma)/2$. The constant $E_{\mathrm{ref}}$ shifts all energies equally and does not affect the gap. The coefficient $B$ mixes states whose boson numbers differ by two, so $A$ alone is not the excitation energy when $B\neq0$.

To find that energy and the corresponding fluctuations, we introduce the two quadratures
\begin{equation}\label{eq:quadrature_ops}
    \hat{X} = (\hat{a} + \hat{a}^\dagger)/\sqrt{2}, 
    \quad \hat{P} = (\hat{a} - \hat{a}^\dagger)/(\ii\sqrt{2}),
\end{equation}
which satisfy $[\hat X,\hat P]=\ii$ and play the roles of position and momentum. Equation~\eqref{eq:hp} gives $\hat S_x\simeq\sqrt{N/2}\,\hat X$ and $\hat S_y\simeq-\sqrt{N/2}\,\hat P$. The same variables therefore describe the two spin components transverse to the mean spin. The quadratic Hamiltonian becomes
\begin{equation}
    \hat{H}_{\mathrm{LMG}} = \frac{K_X}{2}\hat{X}^2 + \frac{K_P}{2}\hat{P}^2 + \mathrm{const.} + \mathcal{O}(N^{-1}),
    \label{eq:stiffness}
\end{equation}
The coefficients $K_X=A+B=2(\omega-g)$ and $K_P=A-B=2(\omega-\gamma g)$ measure the energy costs of the two transverse fluctuations. Both are positive in the stable polarized phase. A nonzero pair coefficient $B$ makes them unequal, while excitation of the mode involves both conjugate directions. The energy of one excitation of this collective mode is, to leading order in $1/N$,
\begin{equation}\label{eq:collective_gap}
    \Delta_{\mathrm{col}} = \sqrt{K_X K_P} = 2\sqrt{(\omega- g)(\omega - \gamma g)}.
\end{equation}
{Within this single-mode description, $\Delta_{\mathrm{col}}$ is the collective excitation gap in the chosen total-spin sector. Section~\ref{sec:deformations} extends the connection to chains with several collective transition energies.}

For $0<g<\omega$, the gap closes as $g$ approaches the critical value $g_c=\omega$. For $0<\gamma<1$, the second formal root at $g=\omega/\gamma$ lies outside the validity of the polarized-state expansion. For $g<0$, both factors in Eq.~\eqref{eq:collective_gap} remain positive and the gap increases monotonically with $|g|/\omega$.

Knowing this excitation energy does not yet tell us how the ground-state fluctuations are shared between the two directions. Writing $\Delta\hat O=\hat O-\langle\hat O\rangle$ for the fluctuation of an observable $\hat O$, the quadrature variances are
\begin{equation}
\langle(\Delta\hat X)^2\rangle=\frac12\sqrt{\frac{K_P}{K_X}},
\qquad
\langle(\Delta\hat P)^2\rangle=\frac12\sqrt{\frac{K_X}{K_P}}.
\label{eq:quadrature_variances}
\end{equation}
The more costly direction has smaller fluctuations, while the conjugate direction has larger fluctuations. Squeezing describes precisely this redistribution, which we can now relate to the collective gap.

\begin{figure}[!tbp]
\centering
\includegraphics[width=\linewidth]{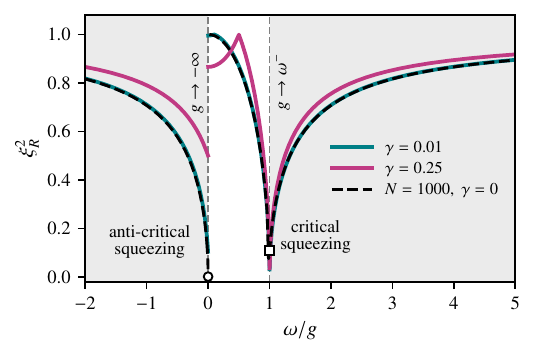}
\caption{Critical and anti-critical squeezing in the anisotropic LMG model.
Solid curves depict the thermodynamic spin-wave results for various anisotropy values $\gamma$. 
The black dashed curve shows a finite-system calculation for $N=1000$ and $\gamma=0$. In the ordered phase, a weak $x$-field of strength $\omega/N^2$ selects one symmetry-broken branch.  
The open square marks the critical point ($\omega/g = 1$), while the open circle marks the even-$N$ anti-critical Ising limit ($\gamma=0$ and $g \to -\infty$), where $\xiR^2 \to 2/(N+2)$. 
In the ordered interval ($0 < \omega/g < 1$), the expansion must follow a selected mean-spin direction (see Appendix~\ref{app:ferromagnetic} for the expansion about a selected minimum).}
\label{fig:lmg_limits}
\end{figure}

\subsection{Critical and anti-critical squeezing}

To express this redistribution in terms of metrological gain, we use the Wineland parameter~{\cite{Wineland1994},}
\begin{equation}
\xiR^2=\frac{N\min_{\bm n_\perp}\langle(\Delta\hat S_{\bm n_\perp})^2\rangle}
{|\langle\hat{\bm S}\rangle|^2},
\label{eq:wineland}
\end{equation}
Here $\bm n_\perp$ is a unit vector perpendicular to the mean spin, and $\hat S_{\bm n_\perp}=\bm n_\perp\cdot\hat{\bm S}$ is the spin component along it. The minimum selects the transverse direction with the smallest variance. A coherent spin state has $\xiR^2=1$, and $\xiR^2<1$ signals metrologically useful squeezing.

For the ground state of the quadratic Hamiltonian, $|\langle\hat S_z\rangle|\simeq N/2$, and the transverse spin variances are $N/2$ times their corresponding quadrature variances. Equations~\eqref{eq:hp} and~\eqref{eq:quadrature_variances} therefore give {the thermodynamic LMG squeezing~\cite{MaWang2009},}
\begin{equation}
\xiR^2
= \sqrt{\frac{\min(K_X,K_P)}{\max(K_X,K_P)}},
    \label{eq:squeezing}
\end{equation}
{The same energy costs $K_X$ and $K_P$ thus determine both $\Delta_{\mathrm{col}}$ and $\xiR^2$, but through different combinations. Their geometric mean sets the gap, whereas their ratio sets the squeezing.} Increasing both costs by the same factor increases the gap without changing either ground-state variance. Changing their relative size changes the squeezing. A gap change alone is therefore not enough to determine the squeezing.

The Ising limit, $\gamma=0$, makes the two possibilities especially clear. Here $K_P=2\omega$ stays fixed, while $K_X=2(\omega-g)$ decreases or increases depending on the sign of $g$. Equations~\eqref{eq:collective_gap} and~\eqref{eq:squeezing} give
\begin{equation}
    \begin{alignedat}{2}
    0 < g < \omega : \qquad & \xiR^2 = \frac{\Delta_{\mathrm{col}}}{2\omega}, \qquad & \Delta_{\mathrm{col}} = 2\sqrt{\omega(\omega - g)}, \\ 
    g < 0 : \qquad & \xiR^2 = \frac{2\omega}{\Delta_{\mathrm{col}}}, \qquad & \Delta_{\mathrm{col}} = 2\sqrt{\omega(\omega + |g|)},
    \end{alignedat}
    \label{eq:critical_anticritical}
\end{equation}
For positive coupling, increasing $g$ reduces the cost of $\hat S_x$ fluctuations. Those fluctuations grow, while the conjugate $\hat S_y$ fluctuations become smaller. Stronger squeezing therefore accompanies a closing gap. Within the thermodynamic quadratic description,
\begin{equation}
\frac{\omega}{g}\to1^+:\qquad
\Delta_{\mathrm{col}}\to0,\qquad
\xiR^2\to0.
\end{equation}
For negative coupling, increasing $|g|$ instead raises the cost of $\hat S_x$ fluctuations. Their variance decreases while the excitation energy increases. The squeezed direction changes from $y$ to $x$, even though the mean spin still points along negative $z$. This is the anti-critical route, with $\xiR^2$ inversely proportional to the collective gap,
\begin{equation}
\frac{\omega}{g}\to0^-:\qquad
\Delta_{\mathrm{col}}\to\infty,\qquad
\xiR^2\to0.
\end{equation}

Although the physical gap changes in opposite directions, the two polarized branches can be written in a common form. For $0\leq\gamma<1$ and $g\neq0$, we compare the gap with the strength $|B|=|g|(1-\gamma)$ of the pair term,
\begin{equation}
\xiR^2
= \frac{\widetilde{\Delta}}
{1+\sqrt{1+\widetilde{\Delta}^2}},
\qquad
\widetilde{\Delta}
\equiv \frac{\Delta_{\mathrm{col}}}{|g|(1-\gamma)}.
\label{eq:lmg_squeezing}
\end{equation}
At $g=0$, the unsqueezed value $\xiR^2=1$ follows by continuity. The normalized gap $\widetilde{\Delta}$ decreases along both Ising routes. Near the positive-coupling critical point, the physical gap closes while $|B|$ remains finite. At large negative coupling, the gap grows as $\sqrt{|g|}$, but $|B|$ grows faster, as $|g|$. In either case, the gap becomes smaller relative to the interaction that creates fluctuation pairs, and squeezing increases.

The anisotropy determines how far this improvement can continue. For $0<\gamma<1$, both energy costs grow at large negative coupling and $K_P/K_X\to\gamma$. Squeezing then saturates at $\xiR^2\to\sqrt{\gamma}$ even though the gap diverges. The Ising limit is special because one energy cost remains fixed, allowing arbitrarily small $\xiR^2$ in the thermodynamic quadratic description. Figure~\ref{fig:lmg_limits} compares these regimes, including the selected ordered state (see Appendix~\ref{app:ferromagnetic} for its local fluctuation theory).

A finite spin rounds these ideal limits. At the critical point $g=\omega$ and fixed $0\leq\gamma<1$, the gap remains nonzero, with both $\Delta_{\mathrm{col}}/\omega$ and $\xiR^2$ scaling as $N^{-1/3}$~\cite{DusuelVidal2004,DusuelVidal2005}. On the anti-critical Ising branch, finite spin length instead gives, at fixed $N$ and $\omega>0$,
\begin{equation}
\lim_{\omega/g\to0^-}\xiR^2=
\begin{cases}
2/(N+2), & N\ \mathrm{even},\\
4N/(N+1)^2, & N\ \mathrm{odd}.
\end{cases}
\label{eq:finite_spin_limits_main}
\end{equation}
{{Both limiting values decrease as $1/N$, in contrast to the $N^{-1/3}$ scaling at criticality. Approaching these limits requires $|g|/(\omega N^2)\gg1$.} The interaction required for a given squeezing is also larger on the gap-opening branch at fixed $\omega$ (see Appendix~\ref{app:finite_spin} for the crossover, coupling requirements, and order of limits).}

The different coefficients reflect the unique lowest state of $\hat S_x^2$ for even $N$ and its $m_x=\pm1/2$ doublet for odd $N$. For even $N$, the ground states of this Hamiltonian family minimize the transverse variance at fixed polarization~\cite{SorensenMolmer2001}. In the even-$N$ strong-coupling limit, both the polarization and the squeezed variance tend to zero, so the corresponding value in Eq.~\eqref{eq:finite_spin_limits_main} is understood as a limit approached from finite coupling. For odd $N$, the best squeezing can occur at finite negative coupling rather than at the asymptotic limit. {The asymptotic state saturates the bound of Ref.~\cite{Caleca2025} for states restricted to the $|m_x=\pm1/2\rangle$ doublet in the maximal-spin sector} (see Appendix~\ref{app:finite_spin} for the even--odd derivation and limiting gaps).

\subsection{Symmetry considerations}
\label{sec:symmetries}

{Both routes toward squeezing rely on the same quadratic pair term. We now ask which symmetries permit pair creation and annihilation, and whether those symmetries alone guarantee squeezing.} Under parity, $\hat a\to-\hat a$, so both $\hat a^\dagger\hat a$ and $\hat a^2+\hat a^{\dagger2}$ remain unchanged. Parity therefore permits a nonzero $B$ on either polarized branch.

The isotropic endpoint provides a useful comparison. At $\gamma=1$, conservation of $\hat S_z$ forces $B=0$, and the two transverse energy costs become equal. The polarized ground state remains unsqueezed, although its collective gap for $g<\omega$,
\begin{equation}
\Delta_{\mathrm{col}}=2(\omega-g)
\end{equation}
can close as $g\to\omega^-$ or grow as $g\to-\infty$. This example shows directly why changing the excitation energy alone does not produce squeezing. The absence of squeezing also holds at finite $N$ away from ground-state degeneracies. The ground state is then an $\hat S_z$ eigenstate, with $\xiR^2\geq1$ whenever its mean spin is nonzero (see Appendix~\ref{app:u1_endpoint} for the exact expression).

The same argument applies to discrete {uniform} rotations about the polarization axis. Under such an axial rotation by angle $\theta$,
\begin{equation}
\hat a\longrightarrow e^{-\ii\theta}\hat a,
\qquad
\hat a^2\longrightarrow e^{-2\ii\theta}\hat a^2.
\label{eq:axial_rotation_pair}
\end{equation}
For $U(1)$ symmetry, $\theta$ can take any value. For $\mathbb Z_n$ symmetry, with integer $n\geq2$, invariance is required under $\theta=2\pi/n$ and its integer multiples. A rotation by $\pi$ leaves $\hat a^2$ unchanged, but a rotation by $2\pi/n$ with $n>2$ does not. These higher axial rotations therefore forbid the pair term when both the Hamiltonian and the polarized reference state preserve the symmetry. The ellipses in Fig.~\ref{fig:mechanism_v16} illustrate the distinction. An ellipse can preserve a rotation by $\pi$ while having different widths along its axes. Preserving a higher axial rotation instead requires equal transverse variances. Among these nontrivial axial rotation symmetries, only $\ZZ$ permits quadratic pair terms about a symmetry-preserving polarized state.

Permission from symmetry is not sufficient, however. The pair coefficient $B$ in Eq.~\eqref{eq:quadratic} can vanish because contributions from different interactions cancel, or an interaction can first contribute beyond quadratic order~\cite{Roscilde2021}. For example, a transverse-field chain with products of four consecutive $x$-spin operators preserves global parity. Around the field-polarized state, each transverse spin is first order in fluctuations, so this interaction first enters at fourth order. Unlike a two-spin Ising interaction, it produces no quadratic pair term. Parity alone therefore does not guarantee the quadratic squeezing mechanism, although higher-order processes can still generate correlations (see Appendix~\ref{app:kbody_symmetry} for the Hamiltonian, sublattice symmetries, and numerical comparison of the two- and four-spin chains).

The assumption about the reference state is equally important. A $U(1)$-symmetric Hamiltonian with degenerate ground states can admit squeezed superpositions that do not preserve $U(1)$. 
{In the odd-sized XX and XXZ models studied in Refs.~\cite{Caleca2025,CalecaLuttinger2025},} a preparation field selects a coherent superposition of ground states whose $\hat S_z$ eigenvalues are $\pm1/2$. An axial rotation changes their relative phase, so the superposition is not invariant and can have a transverse mean spin. After the field is removed, the Hamiltonian regains $U(1)$ symmetry, while the selected state can retain this mean spin. Keeping a finite field instead breaks $U(1)$ explicitly. These states lie outside the symmetry-preserving polarized-state assumption used above. To describe their fluctuations, we must follow the mean-spin direction selected by the state (see Appendix~\ref{app:u1_endpoint} for further details).

\Needspace{10\baselineskip}
\subsection{Explicit and spontaneous symmetry breaking}
\label{sec:breaking}

{When a field or spontaneous symmetry breaking selects the mean-spin direction, the selected state need not preserve the original parity. The question is whether pair fluctuations, and hence the squeezing mechanism, survive around that direction.} In the ordered LMG phase, expanding about either parity-related minimum gives a local quadratic Hamiltonian. Its pair term is generally nonzero, so selecting one minimum does not by itself remove squeezing.

An explicit field separates a change in the mean spin from a change in its fluctuations. A sufficiently weak field can shift the mean without changing the centered variances or excitation energy at quadratic order. A field that tilts the mean spin by a finite angle instead requires a new expansion about that direction. The transverse energy costs can then change, but their geometric mean and ratio still determine the local collective gap and squeezing. Near an instability, even a weak field can require this new expansion (see Appendix~\ref{app:ferromagnetic} for the selected ordered state and the explicit-field derivation).

\FloatBarrier

\section{{Collective excitations and the many-body gap beyond the uniform limit}}
\label{sec:deformations}

Beyond all-to-all interactions, several excitation energies can contribute to the total-spin response. The single-mode relation of Sec.~\ref{sec:lmg} must therefore be extended before we can decide which spectral changes improve squeezing and which also raise the many-body gap. We first establish this connection, then compare XXZ and Ising chains.

\begin{figure*}[t]
\centering
\includegraphics[width=0.98\linewidth]{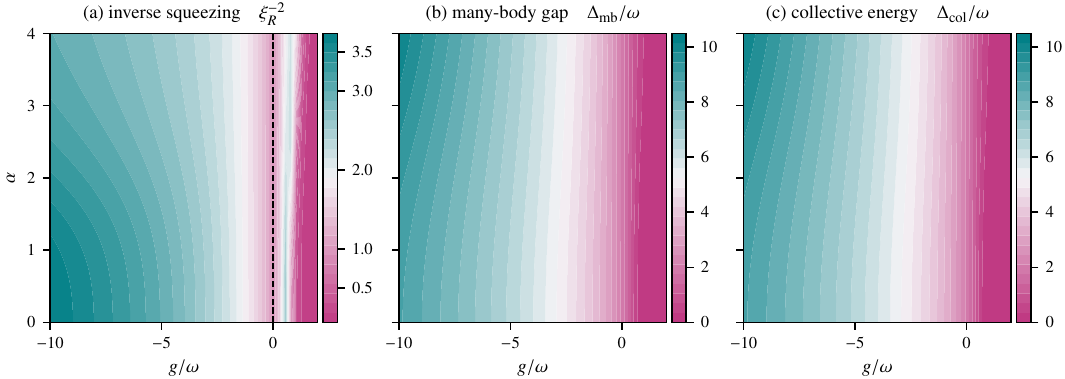}
\caption{Exact results for the XXZ Hamiltonian in Eq.~\eqref{eq:powerlaw} with $N=10$. Panel (a) shows the inverse Wineland parameter $1/\xiR^2$. The black dashed line at $g=0$ separates the minimum-variance axes $\hat S_x$ for $g<0$ and $\hat S_y$ for $g>0$. {At $g=0$, $\xiR^2=1$.} Panel (b) shows the unrestricted many-body gap $\Delta_{\mathrm{mb}}$ and panel (c) the dominant collective excitation energy $\Delta_{\mathrm{col}}$ defined in Eq.~\eqref{eq:collective_gap_definition}. Both are in units of $\omega$. The panels coincide because $W^\perp(E_1)$ is the largest combined transverse weight. Negative coupling increases both $\Delta_{\mathrm{col}}$ and $\Delta_{\mathrm{mb}}$ while improving squeezing. The exact squeezing relation uses $\overline\Delta_x$ and $\Delta_{x,\mathrm{rms}}$ of Eq.~\eqref{eq:ising_response_energies_main}.}
\label{fig:XXZ}
\end{figure*}

\subsection{{The dominant collective excitation}}

The collective spin adds the contributions of all particles with the same sign. An excitation whose transverse contributions cancel in this sum can be absent from the collective-spin response, even if it is the lowest excitation of the system. {This distinction is familiar from dynamical spin structure factors, where each excitation contributes according to its spin-transition strength, not its energy alone~\cite{HohenbergBrinkman1974,Muller1982,CauxMaillet2005}.} Let $|n\rangle$ denote exact eigenstates with energies $E_n$, ordered as $E_0\leq E_1\leq\cdots$, with $|0\rangle$ the ground state. The many-body gap is $\Delta_{\mathrm{mb}}=E_1-E_0$, regardless of how strongly the total spin couples to the first excited state.

{We distinguish the many-body gap from the dominant collective excitation energy $\Delta_{\mathrm{col}}$. In the finite-chain results of Figs.~\ref{fig:XXZ} and~\ref{fig:finite_range_ising}, this is the energy above the ground state of the level carrying the largest combined transverse collective-spin transition strength.} For a nondegenerate ground state with nonzero mean spin and two orthonormal transverse directions $\bm e_1$ and $\bm e_2$, write $\hat S_{\bm e_\mu}=\bm e_\mu\cdot\hat{\bm S}$ and define
\begin{align}
W^\perp(E)&=\sum_{n:E_n=E}\sum_{\mu=1,2}
|\langle n|\hat S_{\bm e_\mu}|0\rangle|^2,\nonumber\\
E_\star&=\underset{E>E_0}{\arg\max}\,W^\perp(E),\nonumber\\
\Delta_{\mathrm{col}}&=E_\star-E_0.
\label{eq:collective_gap_definition}
\end{align}
{The sum within $W^\perp(E)$ includes both transverse directions and all states at energy $E$, making it independent of the eigenvectors chosen within a degenerate subspace. We obtain $\Delta_{\mathrm{col}}=\Delta_{\mathrm{mb}}$ when $W^\perp(E_1)$ is the largest weight. In the single-mode description of Sec.~\ref{sec:lmg}, $\Delta_{\mathrm{col}}$ is the mode's gap. In a chain, however, the largest-weight level can lie above weaker collective transitions and can switch as parameters change. We therefore use $\Delta_{\mathrm{col}}$ as a diagnostic, while relating squeezing to all contributing transitions below (see Appendix~\ref{app:spectral_identity} for the hierarchy of spectral energies and Appendix~\ref{app:ferromagnetic} for local excitations and parity splittings in the ordered phase).}

\subsection{An exact connection to the excitation spectrum}
\label{sec:exact_response}

To go beyond this diagnostic, we need to keep the contributions of all excitations coupled to the collective spin. Both interaction families considered below have a useful property. Their Hamiltonians can be written as
\begin{equation}
\hat H=2\omega\hat S_z+\hat V,
\qquad [\hat V,\hat S_x]=0.
\label{eq:response_class_main}
\end{equation}
Consequently, $[\hat H,\hat S_x]=2\ii\omega\hat S_y$. This relation connects the transition amplitudes of the two transverse spin components without assuming a quadratic Hamiltonian or a fixed total-spin sector. {The connection follows from standard frequency-moment sum rules for spin response~\cite{Muller1982,MonienBedell1992,Stringari1994}. These relate ground-state fluctuations to sums over excitation energies weighted by spin-transition strengths. Related connections between dynamical response and metrological resources have been established for the quantum Fisher information~\cite{Hauke2016}. Here the commutator gives the Wineland parameter directly, without requiring the response to be concentrated in a single excitation.}

{We consider a nondegenerate, parity-preserving ground state with $\langle\hat S_z\rangle<0$, whose mean spin lies along $z$. We denote the excitation energies by $\Delta_n=E_n-E_0$ and their $\hat S_x$ transition strengths by $w_n=|\langle n|\hat S_x|0\rangle|^2$. Each $w_n$ gives the contribution of that excitation to the $\hat S_x$ variance.} {Unlike $W^\perp(E)$, which combines both transverse components at a given energy, $w_n$ contains only the $\hat S_x$ transition strength.} The weighted mean and root-mean-square excitation energies are
\begin{align}
\overline\Delta_x&=\frac{\sum_{n>0}\Delta_n w_n}{\sum_{n>0}w_n},\nonumber\\
\Delta_{x,\mathrm{rms}}^2&=\frac{\sum_{n>0}\Delta_n^2w_n}{\sum_{n>0}w_n}.
\label{eq:ising_response_energies_main}
\end{align}
The commutator then fixes the two transverse Wineland parameters exactly,
\begin{align}
\xi_{R,x}^2&=\frac{N\omega}{|\langle\hat S_z\rangle|\overline\Delta_x},\nonumber\\
\xi_{R,y}^2&=\xi_{R,x}^2
\left(\frac{\Delta_{x,\mathrm{rms}}}{2\omega}\right)^2.
\label{eq:ising_response_ratio_main}
\end{align}
Here $\xi_{R,x}^2$ and $\xi_{R,y}^2$ denote Eq.~\eqref{eq:wineland} evaluated along $x$ and $y$, respectively. {Their symmetrized covariance vanishes, so $\xiR^2=\min(\xi_{R,x}^2,\xi_{R,y}^2)$} (see Appendix~\ref{app:spectral_identity} for the derivation).

{These relations identify what carries over from the collective model. Raising $\overline\Delta_x$ improves $x$ squeezing provided the polarization $|\langle\hat S_z\rangle|$ is retained, while $\Delta_{x,\mathrm{rms}}$ fixes the ratio $\xi_{R,y}^2/\xi_{R,x}^2$. When all nonzero $w_n$ correspond to one excitation energy, $\overline\Delta_x=\Delta_{x,\mathrm{rms}}=\Delta_{\mathrm{col}}$. With $|\langle\hat S_z\rangle|\simeq N/2$, this recovers the direct and inverse single-mode gap dependences in Eq.~\eqref{eq:critical_anticritical}.}

The same relation gives a simple sufficient condition involving the many-body gap. Every contributing excitation obeys $\Delta_n\geq\Delta_{\mathrm{mb}}>0$, hence
\begin{equation}
\xiR^2\leq\xi_{R,x}^2
\leq\frac{N\omega}{|\langle\hat S_z\rangle|\Delta_{\mathrm{mb}}}.
\label{eq:gap_polarization_bound}
\end{equation}
Thus $|\langle\hat S_z\rangle|\Delta_{\mathrm{mb}}>N\omega$ guarantees squeezing for this Hamiltonian class. The condition is sufficient, not necessary. {A low-energy state with $w_n=0$ can lower $\Delta_{\mathrm{mb}}$ and weaken the bound without contributing to either $\overline\Delta_x$ or $\Delta_{x,\mathrm{rms}}$.} This is precisely the distinction we will encounter between XXZ and Ising interactions.

\subsection{XXZ interactions and gap-opening squeezing}

{We begin with an XXZ interaction for which stronger squeezing accompanies both a more energetic dominant collective excitation and a larger many-body gap.} We use the Hamiltonian of Ref.~\cite{Comparin2022}, written with the field along $z$,
\begin{equation}
\hat H=\omega\sum_{i=1}^{N}\hat\sigma_i^z
+\frac{1}{2\mathcal N_N}\sum_{i\neq j}
\frac{g}{r_{ij}^\alpha}
\left(-\hat\sigma_i^x\hat\sigma_j^x
+\hat\sigma_i^y\hat\sigma_j^y
+\hat\sigma_i^z\hat\sigma_j^z\right),
\label{eq:powerlaw}
\end{equation}
where $r_{ij}=\min(|i-j|,N-|i-j|)$ is the distance on a periodic chain, measured in lattice spacings. The XXZ anisotropy axis is now $x$, so the $y$ and $z$ couplings are equal. The exponent $\alpha\geq0$ controls the range, from all-to-all interactions at $\alpha=0$ to nearest-neighbor interactions as $\alpha\to\infty$. The Kac normalization
\begin{equation}
\mathcal N_N=\frac1N\sum_{i\neq j}r_{ij}^{-\alpha}
\label{eq:kac}
\end{equation}
keeps the summed coupling to each spin fixed at a given $g$. The interaction is invariant under collective rotations about $x$, so it commutes with $\hat S_x$. The field breaks that rotation symmetry, but the commutator in Eq.~\eqref{eq:response_class_main} remains exact. The spectral relation and bound therefore apply at every interaction range.

The all-to-all limit connects this model to the collective mechanism. At $\alpha=0$, $\mathcal N_N=N-1$, and $-\hat S_x^2+\hat S_y^2+\hat S_z^2=\hat{\bm S}^2-2\hat S_x^2$ gives
\begin{equation}
\hat H_{\alpha=0}=2\omega\hat S_z+\frac{2g}{N-1}\hat{\bm S}^2-\frac{4g}{N-1}\hat S_x^2+\mathrm{const.}
\label{eq:xxz_alpha_zero}
\end{equation}
Within a fixed total-spin sector, the extra $\hat{\bm S}^2$ term is constant and the remaining Hamiltonian is the Ising LMG model with $g_{\mathrm{LMG}}=2gN/(N-1)$. The extra term nevertheless matters when different sectors are compared. For $g<0$, it favors larger total spin and raises the relative cost of excitations outside $S=N/2$. For the finite all-to-all XXZ systems studied here, the lowest excitation indeed lies within this maximal-spin sector.

At weak coupling, the simultaneous improvement of squeezing and the many-body gap can already be established without a collective approximation. For any fixed finite periodic chain with the normalization in Eq.~\eqref{eq:kac}, perturbation theory gives

\begin{align}
\Delta_{\mathrm{mb}}&=2\omega-2g+O(g^2/\omega),\nonumber\\
\xiR^2&=1-|g|/\omega+O(g^2/\omega^2).
\label{eq:xxz_weak_coupling}
\end{align}

{The linear squeezing correction is a special case of the perturbative ground-state mechanism established in Ref.~\cite{Roscilde2021}.} Negative coupling therefore opens the many-body gap and produces squeezing at weak coupling for every interaction range, including nearest neighbors {(see Appendix~\ref{app:xxz_checks} for the fixed-$N$ derivation of both expressions)}. Distance-dependent interactions can mix states with different total spin, so we next examine the full $N=10$ spectrum. Figure~\ref{fig:XXZ} shows that the negative-coupling branch continues to improve squeezing while increasing $\Delta_{\mathrm{mb}}$ and the dominant collective energy $\Delta_{\mathrm{col}}$. Here $W^\perp(E_1)$ carries most of the total transverse weight $\sum_{E>E_0}W^\perp(E)$, giving $\Delta_{\mathrm{col}}=\Delta_{\mathrm{mb}}$ and identical panels (b) and (c). This concentration connects the finite-chain response to the single-mode picture, while Eq.~\eqref{eq:ising_response_ratio_main} retains the $w_n$ of all excited states.

\subsection{Nearest-neighbor XXZ limit}

{We now ask whether this behavior survives with nearest-neighbor interactions. This limit also allows us to separate the collective mechanism from the other spatial fluctuations analytically.} For a periodic chain with even $N>2$, $\mathcal N_N=2$ and the Hamiltonian is
\begin{align}
\hat H_{\mathrm{XXZ}}^{\mathrm{nn}}={}&\omega\sum_i\hat\sigma_i^z\nonumber\\
&+\frac g2\sum_i\left(-\hat\sigma_i^x\hat\sigma_{i+1}^x
+\hat\sigma_i^y\hat\sigma_{i+1}^y+\hat\sigma_i^z\hat\sigma_{i+1}^z\right).
\label{eq:nn_xxz}
\end{align}
Here the site indices are understood modulo $N$. Taking even $N$ allows a staggered pattern to alternate consistently around the ring.

{We first project onto the fully symmetric sector $S=N/2$ to isolate the collective part~\cite{Plodzien2026} (see Appendix~\ref{app:xxz_checks} for the projector and derivation). With $\hat P_{\mathrm{sym}}$ denoting the projector, the exact matrix elements within this sector give}
\begin{equation}
\hat P_{\mathrm{sym}}\hat H_{\mathrm{XXZ}}^{\mathrm{nn}}\hat P_{\mathrm{sym}}
=\left(2\omega\hat S_z-\frac{4g}{N-1}\hat S_x^2\right)\hat P_{\mathrm{sym}}
+\mathrm{const.}
\label{eq:nn_xxz_projection}
\end{equation}
{The projected Hamiltonian has the Ising LMG form, with $g_{\mathrm{LMG}}=2gN/(N-1)$. To include the spatial fluctuations outside this sector, we separately expand the full-chain Hamiltonian about the polarized state.} Defining the local raising and lowering operators as $\hat\sigma_i^\pm=(\hat\sigma_i^x\pm\ii\hat\sigma_i^y)/2$, the transverse interaction satisfies
\begin{equation}
-\hat\sigma_i^x\hat\sigma_{i+1}^x
+\hat\sigma_i^y\hat\sigma_{i+1}^y
=-2\left(\hat\sigma_i^+\hat\sigma_{i+1}^+
+\hat\sigma_i^-\hat\sigma_{i+1}^-\right).
\label{eq:xxz_pair_only}
\end{equation}
The terms that would move a spin-flip excitation between neighboring sites cancel, leaving pair creation and annihilation. To describe small fluctuations about the negative-$z$-polarized state, we introduce bosonic operators $\hat a_j$, with $\hat a_j^\dagger$ creating a spin flip at site $j$. The local Holstein--Primakoff representation~\cite{HolsteinPrimakoff1940} gives $\hat\sigma_j^z=-1+2\hat a_j^\dagger\hat a_j$. Linearizing the transverse operators gives $\hat\sigma_j^+\simeq\hat a_j^\dagger$ and $\hat\sigma_j^-\simeq\hat a_j$.

Keeping terms up to second order in the bosonic operators, the field contributes $2\omega\sum_j\hat a_j^\dagger\hat a_j$. The $z$--$z$ interaction contributes $-2g\sum_j\hat a_j^\dagger\hat a_j$ because each site belongs to two bonds, while its product of two local occupations is quartic and is neglected. Equation~\eqref{eq:xxz_pair_only} gives the pair term $-g\sum_j(\hat a_j\hat a_{j+1}+\hat a_j^\dagger\hat a_{j+1}^\dagger)$. In terms of the Fourier combinations $\hat a_k=N^{-1/2}\sum_{j=1}^{N}e^{-\ii kj}\hat a_j$, the nearest-neighbor pair term acquires a factor $\cos k$ after combining opposite momenta. The allowed momenta are $k=2\pi m/N$, with $m=0,\ldots,N-1$. The resulting quadratic Hamiltonian and its excitation energy from bosonic Bogoliubov diagonalization~\cite{Colpa1978} are
\begin{align}
\hat H_{\mathrm{XXZ}}^{(2)}={}&\sum_k(2\omega-2g)\hat a_k^\dagger\hat a_k\nonumber\\
&-g\sum_k\cos k
\left(\hat a_k\hat a_{-k}+\hat a_k^\dagger\hat a_{-k}^\dagger\right),
\label{eq:nn_xxz_bdg}\\
\varepsilon_{\mathrm{XXZ}}^{\mathrm{sw}}(k)={}&
2\sqrt{\omega^2-2\omega g+g^2\sin^2 k},
\nonumber
\end{align}
where the superscript $(2)$ denotes the Hamiltonian truncated at quadratic order in the bosonic fluctuation operators, while $\mathrm{sw}$ labels the corresponding excitation energies obtained within spin-wave theory. The uniform mode has $k=0$ and transverse energy costs $K_X=2\omega-4g$ and $K_P=2\omega$, consistent with Eq.~\eqref{eq:nn_xxz_projection} as $N\to\infty$. Negative coupling raises the cost of $\hat S_x$ fluctuations while leaving the conjugate cost fixed, reproducing the anti-critical LMG mechanism. For $g<0$, Eq.~\eqref{eq:nn_xxz_bdg} also shows that the excitation energy increases with $|g|$ at every momentum. Within this approximation, making the collective mode more energetic does not make another spatial fluctuation less energetic.

{The spin-wave expansion requires small local spin-flip occupation. At strong negative coupling, an exact rotation of every second spin maps the model onto an antiferromagnetic Heisenberg chain in a staggered field. Its known thermodynamic gap grows as $|g|^{1/3}\omega^{2/3}$ with a logarithmic correction~\cite{AffleckOshikawa1999}, confirming gap opening beyond the quadratic approximation (see Appendix~\ref{app:xxz_checks} for the mapping and strong-coupling limit).}

\subsection{Thermal robustness of anti-critical squeezing}
\label{sec:thermal_robustness}

The opening many-body gap suggests a benefit beyond the squeezed ground state itself. A larger gap suppresses the relative thermal population of the lowest excited states. Whether squeezing survives also depends on the spin fluctuations in those states, so we calculate the finite-temperature squeezing directly. Thermal squeezing has been studied near Ising criticality and in XXZ preparation schemes~\cite{FrerotRoscilde2018,Comparin2022}. Here we compare gap-closing and gap-opening branches at nearly equal ground-state squeezing.

For the periodic $N=10$ chain, we evaluate the mean spin and variances in Eq.~\eqref{eq:wineland} using the exact Gibbs state $\hat\rho_T$ at temperature $T$, with Boltzmann constant $k_{\mathrm B}=1$. All eigenstates are included with their Boltzmann weights (see Appendix~\ref{app:numerics} for the thermal state and numerical details).

\begin{figure}[tbp]
\centering
\includegraphics[width=\linewidth]{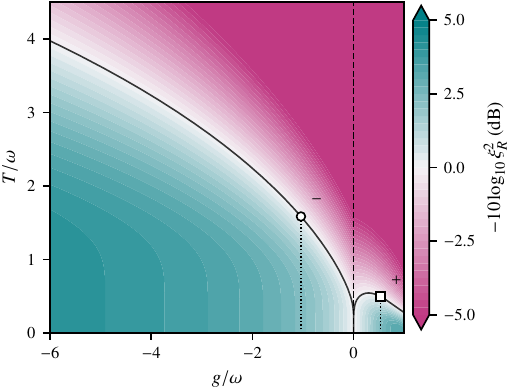}
\caption{Finite-temperature squeezing for the nearest-neighbor limit $\alpha\to\infty$ of Eq.~\eqref{eq:powerlaw} with $N=10$. The color scale shows the metrological gain $-10\log_{10}\xiR^2$ in decibels. Positive values correspond to $\xiR^2<1$ and negative values to $\xiR^2>1$. The dark contour marks $\xiR^2=1$. {The black dashed line marks $g=0$, where the grid includes the analytic result $\xiR^2=\coth^2(\omega/T)>1$ for $T>0$. The black circle ($-$) and square ($+$) mark the squeezing-loss temperatures at $g/\omega=-25/24$ and $8/15$, respectively. The black dotted lines indicate their couplings. These states have nearly equal ground-state squeezing, $\xiR^2\simeq0.58$.} In the squeezed regions, the minimum-variance axis is $\hat S_x$ for $g<0$ and $\hat S_y$ for $g>0$.}
\label{fig:xxz_thermal}
\end{figure}

{To compare thermal robustness without favoring a state that is already more squeezed at zero temperature, we use $g/\omega=8/15$ and $-25/24$, which give nearly equal ground-state squeezing, $\xiR^2\simeq0.58$. We denote the squeezing-loss temperature by $T_*$, defined by $\xiR^2(T_*)=1$. The markers in Fig.~\ref{fig:xxz_thermal} identify this threshold for each state. The gap-closing state loses squeezing at $T_*/\omega\simeq0.49$, whereas the gap-opening state remains squeezed up to $T_*/\omega\simeq1.58$ (see Appendix~\ref{app:numerics} for the size checks).}

{At fixed $\omega$, the negative-coupling state uses a larger interaction. We therefore also compare the states at the same local coupling budget $\Lambda=\omega+3|g|/2$, defined as the sum per site of the absolute Pauli coefficients. Rescaling each Hamiltonian to this budget preserves its ground-state squeezing. The positive- and negative-coupling loss temperatures become $T_*^{(+)}/\Lambda\simeq0.274$ and $T_*^{(-)}/\Lambda\simeq0.618$, respectively. The thermal advantage persists, with little size dependence at these couplings for $N=6,8,10,12$ (see Appendix~\ref{app:numerics} for the normalization and size checks).}

\begin{figure*}[t]
\centering
\includegraphics[width=0.96\linewidth]{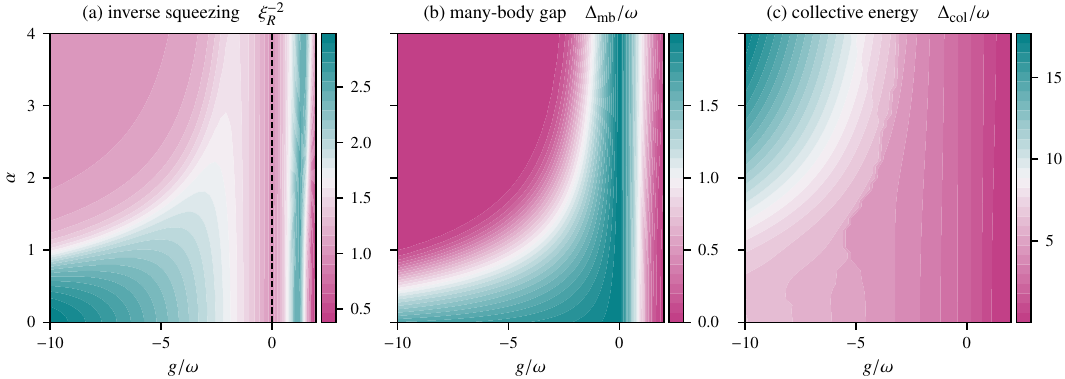}
\caption{Pure power-law Ising results for $N=10$ and periodic boundary conditions. Panel (a) shows $1/\xiR^2$. The black dashed line at $g=0$ separates minimum-variance axes $\hat S_x$ for negative coupling and $\hat S_y$ for positive coupling. {At $g=0$, $\xiR^2=1$.} Panels (b) and (c) show the unrestricted many-body gap $\Delta_{\mathrm{mb}}$ and the dominant collective excitation energy $\Delta_{\mathrm{col}}$ of Eq.~\eqref{eq:collective_gap_definition}, both in units of $\omega$. Their color scales are independent. Negative coupling raises $\Delta_{\mathrm{col}}$ while a lower excitation can keep $\Delta_{\mathrm{mb}}$ small. Squeezing is determined by $\overline\Delta_x$ and $\Delta_{x,\mathrm{rms}}$ in Eq.~\eqref{eq:ising_response_energies_main}. The grid covers $0\leq\alpha\leq4$. The nearest-neighbor limit is analyzed separately in the main text (see Appendix~\ref{app:finite_range_benchmarks} for the projection and excitation dispersions).}
\label{fig:finite_range_ising}
\end{figure*}

\subsection{Power-law Ising interactions}
\label{sec:ising_main}

The Ising chain lets us separate the pair processes that generate squeezing from the interactions that raise the other excitation energies. We retain the same spatial interaction profile as in the XXZ model but keep only the $x$ coupling,
\begin{equation}
\hat H_{\mathrm I}=\omega\sum_i\hat\sigma_i^z-\frac{1}{2\mathcal N_N}\sum_{i\neq j}\frac{g}{r_{ij}^\alpha}\hat\sigma_i^x\hat\sigma_j^x,
\label{eq:powerlaw_ising}
\end{equation}
with the distance and Kac normalization defined in Eqs.~\eqref{eq:powerlaw} and~\eqref{eq:kac}. At $\alpha=0$, total spin is conserved and the Hamiltonian within $S=N/2$ reduces to the Ising LMG model with $g_{\mathrm{LMG}}=gN/(N-1)$. Unlike Eq.~\eqref{eq:xxz_alpha_zero}, however, it has no additional $\hat{\bm S}^2$ term to favor that sector. Conservation of total spin does not require the lowest excitation to remain in the same sector as the ground state. The collective spin operators preserve $S$, so an excitation in another sector can lower the many-body gap without appearing in the collective transition strength.

Figure~\ref{fig:finite_range_ising} shows the resulting separation for $N=10$. Positive coupling squeezes $\hat S_y$ as the dominant collective excitation becomes less energetic. For negative coupling, the squeezed direction is $\hat S_x$ and the collective excitation becomes more energetic. A lower excitation can nevertheless approach the ground state. In the all-to-all model, it belongs to a different total-spin sector. At shorter range, the spins can also fluctuate with different spatial patterns, providing low-energy excitations that are absent from the single collective mode.

The nearest-neighbor chain makes this distinction particularly simple. Neighboring spins fluctuate together in the uniform mode and in opposite directions in the staggered mode. The staggered contributions cancel in the total spin, so a single staggered excitation is not reached by a collective-spin operator. The quadratic dispersion is $\varepsilon_{\mathrm I}^{\mathrm{sw}}(k)=2\sqrt{\omega(\omega-g\cos k)}$. For $g<0$, increasing $|g|$ raises the energy of the uniform $k=0$ mode but lowers that of the staggered $k=\pi$ mode, allowing the many-body gap to decrease even as the collective gap increases.

This spin-wave result predicts the direction of the energy changes, not the exact transition point. The exact Ising dispersion places the transition at $|g|=2\omega$ and gives $\Delta_{\mathrm{mb}}=2\omega-|g|$ in the thermodynamic paramagnetic phase~\cite{LiebSchultzMattis1961,Pfeuty1970,BarouchMcCoy1971} (see Appendix~\ref{app:finite_range_benchmarks} for the projection and both dispersions). The distinction between uniform and staggered fluctuations also explains why uniform squeezing saturates with system size (see Appendix~\ref{app:ising_saturation} for the size dependence of uniform squeezing).

The Ising interaction also commutes with $\hat S_x$, so Eq.~\eqref{eq:ising_response_ratio_main} connects its squeezing to the same weighted energies $\overline\Delta_x$ and $\Delta_{x,\mathrm{rms}}$. Here the total transverse weight $\sum_{E>E_0}W^\perp(E)$ can be spread over several excitation energies rather than concentrated at $E_\star$. A staggered excitation with $w_n=0$ can lower $\Delta_{\mathrm{mb}}$ while contributing to neither $\overline\Delta_x$ nor $\Delta_{x,\mathrm{rms}}$. XXZ and Ising thus share the collective squeezing mechanism, but differ in the energies of the other excitations (see Appendix~\ref{app:spectral_identity} for the spectral comparison).

\section{Conclusions and outlook}
\label{sec:conclusions}

Our results connect ground-state spin squeezing to the excitation spectrum without requiring a closing collective gap. {For a quadratic collective mode}, squeezing reflects the imbalance between the energy costs of conjugate transverse fluctuations, whereas the excitation gap depends on their geometric mean. Stronger squeezing can therefore accompany either increasing or decreasing collective excitation energies. {For the Ising and XXZ interactions considered here, an exact spectral relation connects squeezing to the polarization and to excitation energies weighted by their collective-spin transition strengths.}

An increasing collective excitation energy does not necessarily imply a larger many-body gap. This distinction already arises in the negative-coupling Ising LMG model, where lower-energy excitations outside the maximal-spin sector are inaccessible to collective-spin operators. In the nearest-neighbor Ising chain, a staggered excitation similarly lowers the many-body gap while the uniform mode becomes more energetic. By contrast, in the finite XXZ chains studied here, the dominant collective excitation coincides with the lowest excitation, and stronger squeezing accompanies an increasing many-body gap along the negative-coupling branch. For the nearest-neighbor XXZ chain, this branch also retains squeezing at higher temperatures than the gap-closing branch at comparable ground-state squeezing.

Many studies of critical metrology use $\ZZ$-symmetric Ising, LMG, Rabi, and Dicke systems, where squeezing and a soft mode occur together~\cite{Zanardi2008,Invernizzi2008,Salvatori2014,FrerotRoscilde2018,Garbe2020,Mirkhalaf2021,MontenegroBayat2021,PhysRevLett.130.090802,Salvia2023,Hotter2024,Mihailescu2026}. Our results separate the pair fluctuations that produce Gaussian squeezing from the critical softening that can enhance it. They also give a common perspective on earlier links between ground-state squeezing, entanglement, and phase transitions~\cite{OsborneNielsen2002,Osterloh2002,VidalLatorre2003,AmicoReview2008,MaWang2009}. {Connecting this equilibrium picture to parameter-dependent ground-state sensing~\cite{AndreEtAl2026} and faster preparation~\cite{Rams2018,GietkaSpeeding2022,Young2023,MihailescuGietkaAntiCritical2026,Pedersen2026} requires the corresponding response and ramp matrix elements (see Appendix~\ref{app:spectral_identity} for the metrological distinction and parity selection rules).} Beyond Gaussian squeezing, higher symmetries may instead support useful non-Gaussian states and multipartite resources~\cite{SandboChang2020,WeiTan2024,Lotkov2026,Bruno2026ThreePhoton,JarvisFrain2026,ZhangThreeBody2026}. Finding when such states can be prepared without a closing many-body gap is a natural next question.

\begin{acknowledgments}
K.G. would like to thank Piotr Grochowski, Lewis Ruks and Bubusie for fruitful discussions. This work was supported by the Austrian Science Fund (FWF) through the Cluster of Excellence Quantum Science Austria under Grant No.~10.55776/COE1. A.K. acknowledges the
support of the Polish National Agency for Academic Exchange (NAWA) under the “Polish Returns” programme,
grant no. BPN/PPO/2024/1/00010/U/00001.
\end{acknowledgments}

\begingroup

\section*{AI disclosure}
OpenAI Codex (GPT-6) assisted with explanatory text, literature searches, numerical code, checks of derivations, and figure preparation. Numerical outputs were checked against analytic limits, independent implementations, ground-state eigenvalue residuals, and exact spectral sum rules (see Appendix~\ref{app:numerics} for the verification procedures). The authors take full responsibility for the manuscript.
\endgroup

\begingroup
\section*{Data availability}
The numerical data and computer codes supporting the findings of this work are available from the corresponding author upon reasonable request.
\endgroup

\appendix

\section{Selected mean-spin directions in the LMG model}
\label{app:ferromagnetic}

\subsection{Spontaneous selection in the ferromagnetic phase}

For $g>\omega$ and $0\leq\gamma<1$, the LMG model has two
symmetry-related mean-field minima,
\begin{equation}
\langle\hat{\boldsymbol S}\rangle_{\mathrm{MF}}
=S\left(\pm\sin\theta,0,-\cos\theta\right),
\qquad
\cos\theta=\frac{\omega}{g},
\end{equation}
where $S=N/2$ and $\theta$ is the tilt away from the negative $z$ direction. Taking the thermodynamic limit before removing
an infinitesimal symmetry-breaking field selects one branch.
At finite $N$, a state of definite parity instead combines
contributions from both minima.

To expand about either selected minimum, we rotate the spin operators
so that the new quantization axis follows that minimum,
\begin{equation}
\begin{aligned}
\hat S_x&=\cos\theta\,\hat S'_x
\mp\sin\theta\,\hat S'_z,\\
\hat S_y&=\hat S'_y,\\
\hat S_z&=\pm\sin\theta\,\hat S'_x
+\cos\theta\,\hat S'_z.
\end{aligned}
\label{eq:broken_rotation}
\end{equation}
The selected state is polarized along the negative $z'$ axis.
Applying the Holstein--Primakoff expansion in Eq.~\eqref{eq:hp}
to the primed operators eliminates the linear terms when
$\cos\theta=\omega/g$. The quadratic Hamiltonian is
\begin{equation}
\hat H_{\mathrm{LMG}}^{(2)}
=\frac{K_X}{2}\hat X^2+\frac{K_P}{2}\hat P^2
+\mathrm{const.},
\end{equation}
where the quadratures describe fluctuations in the rotated
frame and
\begin{equation}
K_X=\frac{2(g^2-\omega^2)}{g},
\qquad
K_P=2g(1-\gamma).
\label{eq:broken_stiffnesses}
\end{equation}
Both fluctuation costs are positive in the ferromagnetic phase,
giving the collective gap
\begin{equation}
\Delta_{\mathrm{col}}
=\sqrt{K_XK_P}
=2\sqrt{(g^2-\omega^2)(1-\gamma)}.
\label{eq:broken_gap}
\end{equation}
This gap describes fluctuations about a selected minimum.
At finite $N$, tunneling between the minima produces a
separate parity-doublet splitting.

Equivalently,
\begin{equation}
\hat H_{\mathrm{LMG}}^{(2)}
=\frac{K_X+K_P}{2}\hat a^\dagger\hat a
+\frac{K_X-K_P}{4}
\left(\hat a^2+\hat a^{\dagger2}\right)
+\mathrm{const.},
\label{eq:broken_quadratic}
\end{equation}
with pair coefficient
\begin{equation}
\frac{K_X-K_P}{4}
=\frac{\gamma g^2-\omega^2}{2g}.
\end{equation}
The ground-state covariance in Eq.~\eqref{eq:quadrature_variances} gives the squeezing, while Eq.~\eqref{eq:broken_gap} gives the corresponding local gap,
\begin{equation}
\xiR^2
=\sqrt{\frac{\min(K_X,K_P)}{\max(K_X,K_P)}}.
\label{eq:broken_squeezing_gap}
\end{equation}

For fixed $\gamma<1$, approaching $g=\omega$ from above
softens $K_X$ while $K_P$ remains finite. The collective
gap then closes and $\xiR^2\to0$ within the Gaussian
approximation. For $0<\gamma<1$, the pair term vanishes
at $g=\omega/\sqrt{\gamma}$. Here $K_X=K_P$ and
$\xiR^2=1$, even though the local gap remains finite.
{At leading order in $1/N$, the selected state is then a product of identically oriented spins, connecting the vanishing pair term to ground-state factorization~\cite{GiampaoloAdessoIlluminati2008}.}
The squeezed quadrature changes from $P$ to $X$ across
this point. At larger $g$, squeezing strengthens again
while the collective gap opens, with
$\xiR^2\to\sqrt{1-\gamma}$ as $g\to\infty$.
For $\gamma=0$, the squeezed quadrature remains $P$
and $\xiR^2\to1$ in this limit.

\subsection{Explicit symmetry-breaking fields}

An explicit field allows us to distinguish changes in the mean spin from changes in the fluctuations around it. We first consider a polarized branch and add a field whose strength on each spin vanishes as $N$ grows,
\begin{equation}
\hat H_\eta=\hat H_{\mathrm{LMG}}-\frac{\eta}{\sqrt N}\hat S_x.
\label{eq:source}
\end{equation}
Here $\eta$ is a real field amplitude held fixed as $N\to\infty$. The field breaks parity, but the factor $N^{-1/2}$ keeps a finite linear term for the fluctuation mode while its effect on the mean-spin direction vanishes at fixed coupling away from criticality. Keeping quadratic terms about the polarized state and omitting an additive constant gives
\begin{equation}
\hat H_\eta^{(2)}=A\hat a^\dagger\hat a
+\frac B2(\hat a^2+\hat a^{\dagger2})
-\frac\eta2(\hat a+\hat a^\dagger).
\end{equation}
The linear term shifts the mean value of the bosonic operator. We remove it by writing
\begin{equation}
\hat a=\hat c+\frac{\eta}{2(A+B)}
\end{equation}
where $\hat c$ is a bosonic operator describing fluctuations around the shifted mean. At quadratic order, the shift leaves both transverse variances and the excitation energy unchanged. As long as $|\eta/[2(A+B)]|\ll\sqrt N$ and the small-fluctuation condition remains satisfied, the leading large-$N$ value of $\xiR^2$ is therefore unchanged. When $A+B$ approaches zero, even a weak field can produce a large shift, and we must expand around the new mean-spin direction.

{A field term $-h\hat S_x$, with real $h$ independent of $N$, can instead tilt the mean spin by a finite angle. We first find that direction and choose transverse axes that remove any mixed term in the local quadratic energy. At a stable minimum, the two positive fluctuation costs $K_X$ and $K_P$ give the local gap $\sqrt{K_XK_P}$ and the squeezing $\xiR^2=\sqrt{\min(K_X,K_P)/\max(K_X,K_P)}$, to leading order in $1/N$. Tilting the mean spin therefore does not by itself remove the pair fluctuations around it.}

\section{Finite-spin regularization of the LMG limits}
\label{app:finite_spin}

We now consider how finite spin length modifies the two limiting branches. The positive-coupling critical point is rounded by the leading non-Gaussian correction, while the negative-coupling limit reflects the finite dimension of the spin Hilbert space.

\subsection{Critical rounding and breakdown of the Gaussian expansion}

Near $g_c=\omega$, with $|1-g/\omega|\ll1$, fixed $0\leq\gamma<1$, and large but finite $N$, the quadratic theory becomes singular and the leading finite-spin correction is quartic. The gap in the following equation is the finite-$N$ collective excitation energy, not the vanishing thermodynamic Gaussian gap in Eq.~\eqref{eq:collective_gap}. The continuous unitary transformation analysis of Refs.~\cite{DusuelVidal2004,DusuelVidal2005} gives
\begin{align}
\Delta_{\mathrm{col}}(g_c,N)&\sim C_\Delta(\gamma)\,\omega N^{-1/3},\nonumber\\
\langle(\Delta\hat S_y)^2\rangle&\sim C_y(\gamma)N^{2/3},\nonumber\\
|\langle\hat S_z\rangle|&=\frac N2\left[1+O(N^{-2/3})\right].
\label{eq:critical_finite_scaling}
\end{align}
The Wineland parameter therefore obeys
\begin{equation}
\xiR^2(g_c,N)\sim4C_y(\gamma)N^{-1/3}.
\label{eq:critical_squeezing_scaling}
\end{equation}
{The dimensionless coefficients $C_\Delta(\gamma)$ and $C_y(\gamma)$ depend on anisotropy through the effective critical oscillator.} Matching Eq.~\eqref{eq:critical_squeezing_scaling} to the approach to criticality in Eq.~\eqref{eq:lmg_squeezing} gives the width of the rounded region
\begin{equation}
\left|1-\frac g\omega\right|\sim N^{-2/3}.
\label{eq:critical_window}
\end{equation}

Finite spin length regularizes the anti-critical branch differently. Within the polarized quadratic description, the ground state of the quadratic Hamiltonian contains
\begin{equation}
n_{\mathrm{HP}}=\langle\hat a^\dagger\hat a\rangle
=\frac14\left(\xiR^2+\frac{1}{\xiR^2}-2\right)
\label{eq:depletion}
\end{equation}
Holstein--Primakoff bosons. For $\gamma=0$, $g<0$, and $\xiR^2\ll1$,
\begin{equation}
n_{\mathrm{HP}}\simeq\frac{1}{4\xiR^2}
\simeq\frac14\sqrt{\frac{|g|}{\omega}}.
\end{equation}
The requirement $n_{\mathrm{HP}}\ll N$ therefore becomes
\begin{equation}
\frac{|g|}{\omega}\ll16N^2,
\qquad
\xiR^2\gg\frac{1}{4N}.
\label{eq:hp_validity}
\end{equation}
These inequalities locate the breakdown of the oscillator approximation, not the finite-$N$ limit. A bosonic mode can accommodate arbitrarily many spin-flip fluctuations, whereas the spin-$N/2$ Hilbert space has only $N+1$ states.

\subsection{Strong-coupling limits for even and odd \texorpdfstring{$N$}{N}}

Beyond the Gaussian regime, we treat the transverse field as a perturbation to the dominant interaction. On the anti-critical Ising branch,
\begin{equation}
\hat H=2\omega\hat S_z+\kappa\hat S_x^2,
\qquad \kappa=\frac{2|g|}{N}.
\end{equation}
We take $|g|\to\infty$ at fixed $N$ and fixed $\omega>0$, so that $\kappa\to\infty$. For integer spin, this Hamiltonian also generates states with minimum transverse variance at fixed polarization~\cite{SorensenMolmer2001}.

For even $N$, the spin is integer and the unperturbed ground state is $|m_x=0\rangle$. The field admixes only $|m_x=\pm1\rangle$ at first order
\begin{equation}
|\psi_0\rangle=|0_x\rangle-\frac{2\omega}{\kappa}
\sum_{q=\pm1}|q_x\rangle\langle q_x|\hat S_z|0_x\rangle
+O\!\left(\frac{\omega^2}{\kappa^2}\right).
\end{equation}
Using $\sum_{q=\pm1}|\langle q_x|\hat S_z|0_x\rangle|^2=S(S+1)/2$ gives
\begin{align}
\langle\hat S_z\rangle&=-\frac{2\omega}{\kappa}S(S+1)
+O\!\left(\frac{\omega^3}{\kappa^3}\right),\nonumber\\
\langle(\Delta\hat S_x)^2\rangle&=
\frac{2\omega^2}{\kappa^2}S(S+1)
+O\!\left(\frac{\omega^4}{\kappa^4}\right).
\end{align}
Both quantities vanish, but their combination in the Wineland parameter remains finite. Hence
\begin{equation}
\lim_{|g|/\omega\to\infty}\xiR^2=\frac{2}{N+2}
\qquad (N\ \mathrm{even}).
\label{eq:finite_spin_limit}
\end{equation}
The lowest excited states approach the $m_x=\pm1$ subspace, with $\Delta_{\mathrm{col}}=\kappa+O(\omega^2/\kappa)$. The gap therefore grows without bound.

For odd $N$, the spin is half-integer and the lowest eigenspace of $\hat S_x^2$ is the doublet $|m_x=\pm1/2\rangle$. Let $\hat P_{1/2}$ project onto this doublet and let $\hat\tau_x$ exchange its two basis states. Choosing their relative phase appropriately gives
\begin{equation}
\hat P_{1/2}\hat S_z\hat P_{1/2}
=\frac{S+1/2}{2}\hat\tau_x.
\end{equation}
The transverse field lifts the doublet at first order. Its lower-energy eigenstate has
\begin{equation}
\langle\hat S_z\rangle=-\frac{S+1/2}{2},
\qquad
\langle(\Delta\hat S_x)^2\rangle=\frac14.
\end{equation}
It follows that
\begin{equation}
\lim_{|g|/\omega\to\infty}\xiR^2
=\frac{4N}{(N+1)^2}
\qquad (N\ \mathrm{odd}).
\label{eq:finite_spin_limit_odd}
\end{equation}
The other doublet state becomes the first excited state, with
\begin{equation}
\lim_{|g|/\omega\to\infty}\Delta_{\mathrm{col}}
=2\omega(S+1/2)=\omega(N+1).
\label{eq:finite_spin_gap_odd}
\end{equation}
{After interchanging the $x$ and $z$ axes, this doublet saturates the squeezing bound for states confined to the selected doublet in Ref.~\cite{Caleca2025}. The retained field lifts the degeneracy and produces the limiting gap in Eq.~\eqref{eq:finite_spin_gap_odd}. Both squeezing limits scale as $1/N$, with an odd-$N$ coefficient twice as large. {At finite coupling, admixture of higher $|m_x|$ states can increase the polarization enough to give stronger squeezing than in this doublet limit.} For example, exact diagonalization at $N=21$ gives a minimum $\xiR^2\simeq0.1249$ near $g/\omega=-241$, compared with the asymptotic value $4N/(N+1)^2\simeq0.1736$.} {The same doublet mechanism appears in odd XXZ rings, with the limiting polarization now set by their spatial correlations (see Appendix~\ref{app:odd_xxz} for the comparison).}

\subsection{Coupling requirements and order of limits}

For a target $q=\xiR^2$ with $0<q<1$ in the thermodynamic Ising LMG description, Eq.~\eqref{eq:critical_anticritical} gives the two required couplings,
\begin{equation}
\frac{g_+}{\omega}=1-q^2,
\qquad
\frac{g_-}{\omega}=1-q^{-2}.
\label{eq:lmg_matched_couplings}
\end{equation}
Their magnitudes obey $|g_-|/g_+=q^{-2}$. The opening gap at fixed $\omega$ is therefore obtained with an increasingly larger interaction as the target squeezing improves. Rescaling the entire Hamiltonian leaves its ground-state squeezing unchanged but rescales every physical gap.

The same resource distinction matters for the finite-spin limits. Reaching their asymptotic coefficients requires $|g|/(\omega N^2)\gg1$. At fixed $|g|$, one could satisfy this condition by decreasing $\omega$ with $N$, but the gap would then not grow. For even $N$, it approaches $2|g|/N$. For odd $N$, it approaches $\omega(N+1)$, which also vanishes along such a sequence. The $1/N$ squeezing limits consequently do not imply a size-independent gap under a fixed interaction budget.

\section[The finite-spin U(1) endpoint]{The finite-spin \texorpdfstring{$U(1)$}{U(1)} endpoint}
\label{app:u1_endpoint}

The $U(1)$ endpoint provides a finite-spin check of the symmetry argument. Its interaction conserves $\hat S_z$ and cannot create fluctuation pairs around a polarized state. In the maximal-spin sector,
\begin{align}
\hat H_{U(1)}
&=2\omega\hat S_z-\frac{2g}{N}
(\hat S_x^2+\hat S_y^2)
\nonumber\\
&=2\omega\hat S_z+\frac{2g}{N}\hat S_z^2+\mathrm{const.}
\label{eq:u1_finite_hamiltonian}
\end{align}
 The eigenstates are $|S,m\rangle$. 
More generally, for $m\neq0$,
\begin{align}
\langle(\Delta\hat S_x)^2\rangle
=\langle(\Delta\hat S_y)^2\rangle
&=\frac12\left[S(S+1)-m^2\right],\nonumber\\
\xiR^2&=\frac{N[S(S+1)-m^2]}{2m^2}\geq1,
\label{eq:u1_finite_squeezing}
\end{align}
where the inequality assumes $S=N/2$ and equality holds only for $|m|=S$. At $m=0$, the mean spin vanishes and the Wineland parameter is undefined. Away from level degeneracies, the ground state is a definite-$m$ state and is not Wineland squeezed. At a degeneracy, a coherent superposition can instead break the continuous symmetry.

We now turn to the zero-field models of Refs.~\cite{Caleca2025,CalecaLuttinger2025}, rather than Eq.~\eqref{eq:u1_finite_hamiltonian} at fixed $\omega>0$. In these comparison models, the longitudinal term $2\omega\hat S_z$ is absent. Their Hamiltonian conserves $\hat S_z$ and the $x$-axis parity $\hat P^x=\prod_i\hat\sigma_i^x$, which reverses $\hat S_z$ and differs from $\hat\Pi_z$. The ground doublet has $\hat S_z=\pm1/2$. An additional preparation field $-h\hat S_x$ breaks $U(1)$ but preserves $\hat P^x$, so a slow ramp can select the equal superposition
\begin{equation}
|\psi_+\rangle
=\frac{|\Psi_{1/2}\rangle+|\Psi_{-1/2}\rangle}{\sqrt2}.
\label{eq:u1_selected_doublet}
\end{equation}
Here $|\Psi_m\rangle$ denotes the ground state in the sector $\hat S_z=m$. For finite-range interactions, these states need not be individual Dicke states because total spin is not conserved. A rotation by angle $\phi$ about $z$ gives
\begin{equation}
e^{-\ii\phi\hat S_z}|\psi_+\rangle
=\frac{e^{-\ii\phi/2}|\Psi_{1/2}\rangle
+e^{\ii\phi/2}|\Psi_{-1/2}\rangle}{\sqrt2}.
\end{equation}
Each definite-$m$ state acquires only an overall phase, whereas their superposition acquires a relative phase and changes its transverse orientation. {The selected state $|\psi_+\rangle$ can therefore have $\langle\hat S_x\rangle\neq0$ and remains a ground state after the field is removed because the doublet is degenerate. At the zero-field endpoint, $|\psi_+\rangle$ contains only the two sectors $\hat S_z=\pm1/2$, giving $\langle(\Delta\hat S_z)^2\rangle=1/4$. The $x$-axis parity $\hat P^x$ preserved during the ramp selects the relative sign of these two components.} The restriction to the ground doublet fixes the variance, not parity alone. The resulting squeezing scales as $1/N$ with long-range order and also improves with system size in the gapless Luttinger-liquid models of Ref.~\cite{CalecaLuttinger2025}.

The field need not be removed to obtain squeezing. A small finite field can give stronger squeezing for odd $N$ and also permits ground-state squeezing for even $N$~\cite{Caleca2025,CalecaLuttinger2025}. At finite preparation field, the Hamiltonian breaks $U(1)$ explicitly, so $\hat S_z$ is not conserved and other magnetization sectors can contribute. The restriction on pair terms concerns fluctuations about a polarized reference state that preserves the {uniform} axial symmetry. It does not exclude either form of symmetry selection.


\section{Higher-order interactions and the limits of parity}
\label{app:kbody_symmetry}

\begin{figure}[!tbp]
\centering
\includegraphics[width=\linewidth]{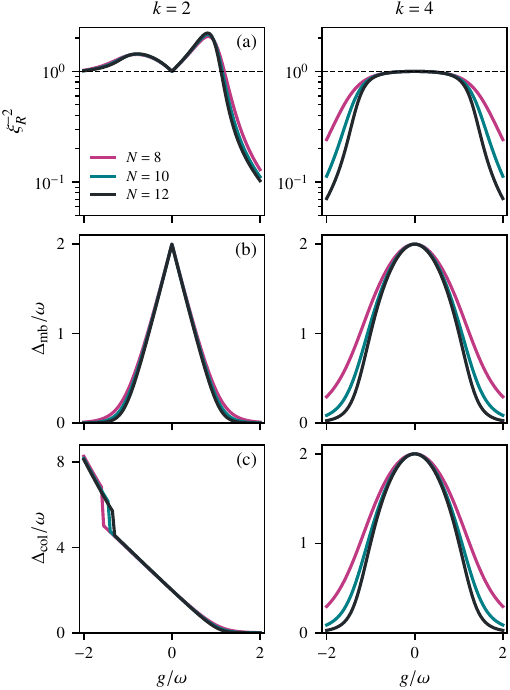}
\caption{Exact ground-state results for the open chain in Eq.~\eqref{eq:kbody}. Columns compare $k=2$ and $k=4$ for $N=8$, $10$, and $12$, with $\omega=1$. {Row (a) shows $1/\xiR^2$, with the black dashed line marking the unsqueezed value. Row (b) shows the unrestricted many-body gap $\Delta_{\mathrm{mb}}/\omega$. Row (c) shows the dominant collective excitation energy $\Delta_{\mathrm{col}}/\omega$, defined from $W^\perp(E)$ in Eq.~\eqref{eq:collective_gap_definition}. The columns in row (c) have different vertical scales. Jumps mark changes of the largest-weight transition. For $k=2$, negative coupling separates the collective excitation energy from the decreasing many-body gap. For $k=4$, both energies are symmetric under $g\to-g$ and closely follow one another on this grid, while the ground state remains unsqueezed. Both interactions preserve global parity, but only $k=2$ generates quadratic pair terms around the polarized state.}}
\label{fig:kbody}
\end{figure}

To compare quadratic and higher-order interactions, we use a multispin Ising chain~\cite{PensonJullienPfeuty1982}. For an open chain of $N$ spins, its Hamiltonian is
\begin{align}
\hat H_k={}&\omega\sum_{i=1}^{N}\hat\sigma_i^z
-g\sum_{i=1}^{N-k+1}
\prod_{j=0}^{k-1}\hat\sigma_{i+j}^x.
\label{eq:kbody}
\end{align}
Here $k$ is the number of consecutive spins in each interaction term, with $2\leq k\leq N$. The $k=2$ case is the transverse-field Ising chain. The $k=4$ chain provides a counterexample to identifying parity with squeezing. Like $k=2$, it preserves global spin parity, but its interaction first couples fluctuations at fourth order. We give the symmetry generators explicitly and compare the exact ground states in Fig.~\ref{fig:kbody}.

\subsection{Sublattice symmetry and the polarized expansion}

For an open chain with $N\geq k$, group the sites by their index modulo $k$. Each contiguous interaction block contains exactly one site from every sublattice. Define the sublattice parities and a set of independent generators by
\begin{align}
\hat P_m&=\prod_{j\equiv m\ ({\rm mod}\ k)}\hat\sigma_j^z,
\qquad m=0,\ldots,k-1,\nonumber\\
\hat U_m&=\hat P_{m-1}\hat P_m,
\qquad m=1,\ldots,k-1.
\label{eq:kbody_generators}
\end{align}
Conjugation by one sublattice parity changes the sign of every interaction block, whereas a product of two sublattice parities leaves it unchanged. The field term commutes with every $\hat P_m$. The $k-1$ commuting generators $\hat U_m$ therefore give a $(\mathbb Z_2)^{k-1}$ symmetry of the full Hamiltonian. The global spin parity defined in the main text is
\begin{equation}
\hat\Pi_z=(-1)^N\prod_{m=0}^{k-1}\hat P_m.
\end{equation}
It is conserved for even $k$, including $k=4$, but reverses the interaction sign for odd $k$. Moreover,
\begin{equation}
\hat P_m\hat H_k(g)\hat P_m=\hat H_k(-g),
\label{eq:kbody_sign_rotation}
\end{equation}
so the spectra and many-body gaps are identical for the two coupling signs. The total transverse spin changes under this transformation, so equal gaps do not imply equal squeezing.

{For $k=4$, the sublattice symmetries make transverse-spin transitions on different sublattices belong to different symmetry sectors. The rotation $\hat P_m$ reverses the sign of the contribution from one sublattice without changing its transition weight. Consequently, $W^\perp(E)$ and $\Delta_{\mathrm{col}}$ are identical for the two coupling signs. For $k=2$, contributions from the two sublattices can interfere, so their collective transition weights need not be symmetric under $g\to-g$.}

Around the negative-$z$-polarized reference state, we use the local Holstein--Primakoff representation $\hat\sigma_i^z=-1+2\hat a_i^\dagger\hat a_i$ and retain only the linear part of the transverse operator, $\hat\sigma_i^x\simeq\hat a_i+\hat a_i^\dagger$. This approximation neglects nonlinear corrections associated with the finite local spin length. Keeping terms up to second order in the bosonic operators gives, for $k=2$,
\begin{align}
\hat H_2^{(2)}={}&2\omega\sum_i\hat a_i^\dagger\hat a_i\nonumber\\
&-g\sum_{i=1}^{N-1}(\hat a_i+\hat a_i^\dagger)
(\hat a_{i+1}+\hat a_{i+1}^\dagger)+\mathrm{const.}
\label{eq:kbody_quadratic_two}
\end{align}
The interaction contains both hopping and pair creation or annihilation. For $k>2$, each block instead starts at order $k$ in fluctuations, leaving
\begin{equation}
\hat H_k^{(2)}=2\omega\sum_i\hat a_i^\dagger\hat a_i
+\mathrm{const.},\qquad k>2.
\label{eq:kbody_quadratic_higher}
\end{equation}
In particular, the $k=4$ interaction enters at quartic rather than quadratic order. Its quadratic vacuum is unsqueezed, even though the exact interaction changes the ground state and excitation spectrum through higher-order fluctuations.

\subsection{Numerical comparison}

{For Fig.~\ref{fig:kbody}, we use Eq.~\eqref{eq:kbody} with $N=8$, $10$, and $12$, set $\omega=1$, and sample $81$ values of $g/\omega\in[-2,2]$. A sparse eigensolver with tolerance $10^{-12}$ resolves the two lowest energies in every sector of the conserved sublattice symmetries. Comparing these sectors gives the global ground state and $\Delta_{\mathrm{mb}}$. We calculate the Wineland parameter from the smallest transverse covariance eigenvalue, as in Eq.~\eqref{eq:wineland}.}

{To obtain $\Delta_{\mathrm{col}}$, we diagonalize every symmetry sector reached by $\hat S_x|0\rangle$ or $\hat S_y|0\rangle$ and retain its full response. For $k=2$, reflection symmetry further reduces this calculation without discarding collective weight. We sum both transverse components and all exactly degenerate levels before selecting the largest $W^\perp(E)$. Since the interaction commutes with $\hat S_x$, the same exact spectral identities apply here (see Appendix~\ref{app:spectral_identity} for their derivation). The exact sum rules for the $w_n$ moments and total $W^\perp(E)$ weight are satisfied within $2\times10^{-12}$ in units with $\omega=1$. Independent full-Hilbert-space diagonalization for $N=8$ verifies the collective energies at selected couplings, while a full-space Krylov calculation checks the squeezing and many-body gaps across the grid.}

\section{Exact spectral relation for XXZ and Ising interactions}
\label{app:spectral_identity}

{We first note a sum rule for the combined transverse weights $W^\perp(E)$ defined in Eq.~\eqref{eq:collective_gap_definition}.} Completeness gives
\begin{equation}
\sum_{E>E_0}W^\perp(E)
=\sum_{\mu=1,2}
\langle 0|(\Delta\hat S_{\bm e_\mu})^2|0\rangle.
\label{eq:collective_sum_rule}
\end{equation}
We assume a nondegenerate ground state and sum once over each distinct excitation energy. {The fraction $W^\perp(E_\star)/\sum_{E>E_0}W^\perp(E)$ measures how much of the combined transverse response comes from the selected level. It differs from a fraction formed using the $\hat S_x$ weights $w_n$ alone.}

\subsection{Response-moment identity}

Frequency-moment sum rules give the exact connection between squeezing and the collective-spin response used in Sec.~\ref{sec:exact_response}~\cite{Muller1982,MonienBedell1992}. Consider
\begin{equation}
\hat H=2\omega\hat S_z+\hat V,
\qquad [\hat V,\hat S_x]=0,
\label{eq:spectral_general_H}
\end{equation}
where $\hat V$ may contain Ising interactions or the XXZ interactions of Eq.~\eqref{eq:powerlaw}. For the latter, each $x$--$x$ term commutes with $\hat S_x$, as does each sum $\hat\sigma_i^y\hat\sigma_j^y+\hat\sigma_i^z\hat\sigma_j^z$. The spatial coefficients are arbitrary. We take a nondegenerate ground state $|0\rangle$ with $\langle\hat S_z\rangle<0$. No assumption about permutation symmetry or Gaussian fluctuations is required.

Equation~\eqref{eq:spectral_general_H} implies
\begin{equation}
[\hat H,\hat S_x]=2\ii\omega\hat S_y.
\label{eq:spectral_commutator}
\end{equation}
Taking its matrix element between the ground state and an excited state gives
\begin{equation}
\Delta_n\langle n|\hat S_x|0\rangle
=2\ii\omega\langle n|\hat S_y|0\rangle,
\qquad \Delta_n=E_n-E_0.
\label{eq:spectral_matrix_element}
\end{equation}
The ground-state expectation value of Eq.~\eqref{eq:spectral_commutator} also gives $\langle\hat S_y\rangle=0$. {We use the $\hat S_x$ transition weights $w_n$ and the corresponding energies $\overline\Delta_x$ and $\Delta_{x,\mathrm{rms}}$ introduced with Eq.~\eqref{eq:ising_response_energies_main}.} Completeness and Eq.~\eqref{eq:spectral_matrix_element} give the exact ratio
\begin{equation}
\frac{\langle(\Delta\hat S_y)^2\rangle}
{\langle(\Delta\hat S_x)^2\rangle}
=\left(\frac{\Delta_{x,\mathrm{rms}}}{2\omega}\right)^2.
\label{eq:spectral_ratio_appendix}
\end{equation}

{The first moment $\sum_{n>0}\Delta_n w_n$ follows from the double-commutator sum rule}
\begin{align}
\sum_{n>0}\Delta_n|\langle n|\hat S_x|0\rangle|^2
&=\frac12\langle0|[\hat S_x,[\hat H,\hat S_x]]|0\rangle
\nonumber\\
&=-\omega\langle\hat S_z\rangle
=\omega|\langle\hat S_z\rangle|.
\label{eq:first_moment_sum_rule}
\end{align}
Consequently,
\begin{equation}
\overline{\Delta}_x
=\frac{\omega|\langle\hat S_z\rangle|}
{\langle(\Delta\hat S_x)^2\rangle}.
\label{eq:spectral_centroid}
\end{equation}
The same matrix-element relation fixes the symmetrized covariance
\begin{equation}
\frac12\langle\{\Delta\hat S_x,\Delta\hat S_y\}\rangle=0.
\label{eq:spectral_covariance}
\end{equation}
Thus $\hat S_x$ and $\hat S_y$ are principal covariance axes within their plane. For a parity-symmetric ground state, the mean spin lies along $z$ and both axes are transverse. If a parity-breaking term still commutes with $\hat S_x$, Eqs.~\eqref{eq:spectral_ratio_appendix} and~\eqref{eq:spectral_centroid} remain exact. However, $\hat S_x$ need not remain perpendicular to the tilted mean spin.

For a mean spin along $z$, Eqs.~\eqref{eq:spectral_ratio_appendix} and \eqref{eq:spectral_centroid} give the two axis-resolved Wineland parameters directly
\begin{align}
\xi_{R,x}^2
&=\frac{N}{2|\langle\hat S_z\rangle|}
\frac{2\omega}{\overline{\Delta}_x},
\nonumber\\
\xi_{R,y}^2
&=\frac{N}{2|\langle\hat S_z\rangle|}
\frac{\Delta_{x,\mathrm{rms}}^2}
{2\omega\overline{\Delta}_x}.
\label{eq:exact_spectral_reconstruction}
\end{align}
{Equation~\eqref{eq:exact_spectral_reconstruction} holds in the full Hilbert space.}

To compare these energies with the many-body gap, define the lowest excitation coupled to the ground state by $\hat S_x$ as
\begin{equation}
\Delta_x=\min_{\{n>0\,|\,\langle n|\hat S_x|0\rangle\neq0\}}\Delta_n.
\label{eq:spectral_visible_gap}
\end{equation}
The four energies obey
\begin{equation}
\Delta_{\mathrm{mb}}\leq\Delta_x
\leq\overline{\Delta}_x\leq\Delta_{x,\mathrm{rms}}.
\label{eq:spectral_energy_hierarchy}
\end{equation}
{When all nonzero $w_n$ correspond to one energy, $\Delta_x=\overline\Delta_x=\Delta_{x,\mathrm{rms}}=\Delta_{\mathrm{col}}$, giving the single-mode reduction discussed in Sec.~\ref{sec:exact_response}. With several contributing energies, $\Delta_x$ is the lowest coupled excitation energy, $\overline\Delta_x$ and $\Delta_{x,\mathrm{rms}}$ are the weighted averages, and $\Delta_{\mathrm{mb}}$ is the unrestricted many-body gap. Combining this hierarchy with Eq.~\eqref{eq:exact_spectral_reconstruction} gives the bound in Eq.~\eqref{eq:gap_polarization_bound}. Even when $\Delta_{\mathrm{col}}=\Delta_{\mathrm{mb}}$, transitions at higher energies can keep $\overline\Delta_x>\Delta_{\mathrm{mb}}$ and make the bound strict.}

{For an explicit comparison, consider $N=10$. In the nearest-neighbor XXZ chain at $g/\omega=-25/24$, $W^\perp(E_\star)$ accounts for $98.11\%$ of $\sum_{E>E_0}W^\perp(E)$, with}
\begin{align}
\Delta_{\mathrm{mb}}/\omega=\Delta_{\mathrm{col}}/\omega&\simeq3.8057,\nonumber\\
\overline\Delta_x/\omega&\simeq3.8274,\nonumber\\
\Delta_{x,\mathrm{rms}}/\omega&\simeq3.8420.
\end{align}
The single-mode approximation is accurate here, although not exact. {At $\alpha=2.64$ and $g/\omega=-3.76$ in the Ising grid, $W^\perp(E_\star)$ instead accounts for $34.52\%$ of the same total transverse weight, and}
\begin{align}
\Delta_{\mathrm{mb}}/\omega&\simeq0.03862,\nonumber\\
\Delta_{\mathrm{col}}/\omega&\simeq5.1328,\nonumber\\
\overline\Delta_x/\omega&\simeq5.6563,\nonumber\\
\Delta_{x,\mathrm{rms}}/\omega&\simeq5.6811.
\end{align}
With $|\langle\hat S_z\rangle|\simeq2.2600$, the exact moments give $\xiR^2\simeq0.7823$. {Replacing $\overline\Delta_x$ by $\Delta_{\mathrm{col}}$ would not reproduce this value.} Full-spectrum diagonalization verifies Eq.~\eqref{eq:exact_spectral_reconstruction} at both points to numerical precision.

\subsection{{Metrological use and adiabatic preparation}}

{The Wineland parameter quantifies sensitivity to a collective rotation of a prepared state. In ground-state sensing, the unknown parameter instead enters the Hamiltonian and changes its ground state~\cite{AndreEtAl2026,Abiuso2025}. The relevant transition matrix elements involve the derivative of the Hamiltonian with respect to the parameter. A collective-spin response and a parameter-estimation response can therefore weight the same spectrum differently. The broader trade-off between precision and equilibration time has also been studied for probes with two-body interactions~\cite{Puig2025}.}

The excitation spectrum relevant to a slow ramp also depends on the operator that changes the Hamiltonian. For an interaction ramp $g(t)$, first-order adiabatic perturbation theory contains the dimensionless ratios~{\cite{GietkaMetzKellerLi2021}}
\begin{equation}
\frac{|\dot g\,\langle n|\partial_g\hat H|0\rangle|}
{(E_n-E_0)^2}.
\label{eq:ramp_matrix_elements}
\end{equation}
{For a parity-preserving ramp, $\partial_g\hat H$ connects the ground state to excitations of the same parity, whereas $\hat S_x$ and $\hat S_y$ connect opposite parities. Evaluating preparation times therefore requires the transitions allowed by the ramp operator and their evolution along the chosen path.}

\Needspace{10\baselineskip}
\section{XXZ checks beyond spin-wave theory}
\label{app:xxz_checks}

{Here we derive the weak-coupling squeezing and gap trends and construct the projection onto the fully symmetric subspace. We then treat the nearest-neighbor strong-coupling limit, using an exact mapping for even rings and a ground-state doublet for the odd rings checked below.}

\subsection{Weak coupling at arbitrary interaction range}

Write Eq.~\eqref{eq:powerlaw} as a sum over unordered pairs with $J_{ij}=g r_{ij}^{-\alpha}/\mathcal N_N$. Translation invariance and the Kac normalization give $\sum_{j\neq i}J_{ij}=g$. At $g=0$, the ground state $|{\downarrow}\cdots{\downarrow}\rangle$ is fully polarized and the one-spin-flip excitations cost $2\omega$.

The interaction $-\hat\sigma_i^x\hat\sigma_j^x+\hat\sigma_i^y\hat\sigma_j^y$ only creates or removes pairs. Its matrix elements within the one-spin-flip subspace vanish, so there is no first-order hopping or splitting within that manifold. The $z$--$z$ term changes the energy of a flipped spin relative to the ground state by $-2\sum_{j\neq i}J_{ij}=-2g$. Thus
\begin{equation}
\Delta_{\mathrm{mb}}=2\omega-2g+O(g^2/\omega)
\end{equation}
at any fixed finite $N$ and sufficiently small $|g|/\omega$.

Let $|ij\rangle$ denote the polarized product state with spins $i$ and $j$ flipped. Its unperturbed excitation energy is $4\omega$, and the first-order ground state is
\begin{equation}
|0\rangle=|{\downarrow}\cdots{\downarrow}\rangle
+\sum_{i<j}\frac{J_{ij}}{2\omega}|ij\rangle
+O(g^2/\omega^2).
\end{equation}
Since $\sum_{i<j}J_{ij}=Ng/2$, the resulting moments are
\begin{align}
\langle\hat S_z\rangle&=-N/2+O(g^2/\omega^2),\nonumber\\
\langle(\Delta\hat S_x)^2\rangle&=\frac N4(1+g/\omega)+O(g^2/\omega^2),\nonumber\\
\langle(\Delta\hat S_y)^2\rangle&=\frac N4(1-g/\omega)+O(g^2/\omega^2).
\end{align}
Minimizing the transverse variance gives Eq.~\eqref{eq:xxz_weak_coupling}. The remainders here are taken at fixed $N$ and fixed spatial profile. These expressions establish the weak-coupling trend, not a uniform error estimate as $N\to\infty$.

{\subsection{Projection onto the fully symmetric subspace}

The projector in Eq.~\eqref{eq:nn_xxz_projection} keeps the states with maximal total spin $S=N/2$. Let $|S,m\rangle$ denote normalized Dicke states with $\hat S_z$ eigenvalue $m$~\cite{Dicke1954}. Then
\begin{equation}
\hat P_{\mathrm{sym}}
=\sum_{m=-S}^{S}|S,m\rangle\langle S,m|,
\qquad S=N/2.
\label{eq:symmetric_projector}
\end{equation}
Every state in this subspace is unchanged by exchanging any two spins. Consequently, all distinct pairs have the same matrix elements between symmetric states. Using $\sum_{i\neq j}\hat\sigma_i^\mu\hat\sigma_j^\mu=4\hat S_\mu^2-N$ gives
\begin{equation}
\hat P_{\mathrm{sym}}\hat\sigma_i^\mu\hat\sigma_j^\mu\hat P_{\mathrm{sym}}
=\frac{4\hat S_\mu^2-N}{N(N-1)}\hat P_{\mathrm{sym}},
\quad i\neq j,
\label{eq:symmetric_pair_projection}
\end{equation}
where $\mu=x,y,z$. For a periodic chain with $N>2$, summing its $N$ nearest-neighbor bonds therefore yields
\begin{equation}
\hat P_{\mathrm{sym}}\sum_i\hat\sigma_i^\mu\hat\sigma_{i+1}^\mu\hat P_{\mathrm{sym}}
=\frac{4\hat S_\mu^2-N}{N-1}\hat P_{\mathrm{sym}}.
\label{eq:symmetric_bond_projection}
\end{equation}
Substituting this identity into Eq.~\eqref{eq:nn_xxz} gives an interaction proportional to $4(\hat{\boldsymbol S}^{\,2}-2\hat S_x^2)-N$. Within the symmetric subspace, $\hat{\boldsymbol S}^{\,2}=S(S+1)=N(N+2)/4$, so
\begin{align}
\hat P_{\mathrm{sym}}\hat H_{\mathrm{XXZ}}^{\mathrm{nn}}\hat P_{\mathrm{sym}}
={}&\bigg[2\omega\hat S_z-\frac{4g}{N-1}\hat S_x^2\nonumber\\
&\quad+\frac{gN(N+1)}{2(N-1)}\bigg]\hat P_{\mathrm{sym}}.
\label{eq:xxz_projection_full}
\end{align}
Dropping the additive constant gives Eq.~\eqref{eq:nn_xxz_projection}. Applying Eq.~\eqref{eq:symmetric_bond_projection} to the Ising interaction likewise gives Eq.~\eqref{eq:nn_ising_projection}, with constant $gN/[2(N-1)]$. These projection identities hold for even and odd $N$. They determine the matrix elements between symmetric states. Nearest-neighbor interactions can also couple these states to the rest of the Hilbert space, so the projected Hamiltonian does not by itself determine the full excitation spectrum.}

\subsection{Mapping the nearest-neighbor negative-coupling branch}

For an even periodic chain, let $\hat U=\prod_{j\,\mathrm{even}}\hat\sigma_j^x$ and $\epsilon_j=(-1)^{j+1}$. The rotation leaves $\hat\sigma_j^x$ unchanged and reverses $\hat\sigma_j^{y,z}$ on every even site. Each nearest-neighbor bond contains exactly one rotated spin. With $g<0$, Eq.~\eqref{eq:nn_xxz} therefore becomes
\begin{equation}
\hat U\hat H_{\mathrm{XXZ}}^{\mathrm{nn}}\hat U^\dagger
=\frac{|g|}{2}\sum_j\hat{\bm\sigma}_j\cdot\hat{\bm\sigma}_{j+1}
+\omega\sum_j\epsilon_j\hat\sigma_j^z.
\label{eq:xxz_staggered_mapping}
\end{equation}
This is the antiferromagnetic Heisenberg chain in a staggered field. In spin-$1/2$ notation $\hat{\bm s}_j=\hat{\bm\sigma}_j/2$, its exchange is $J=2|g|$ and staggered-field magnitude is $h_s=2\omega$. Taking the thermodynamic limit before $h_s/J\to0$, the known low-field result gives~\cite{AffleckOshikawa1999}
\begin{equation}
\Delta_{\mathrm{mb}}
\propto |g|^{1/3}\omega^{2/3}
\left[\ln(|g|/\omega)\right]^{1/6}.
\label{eq:xxz_exact_asymptotic_gap}
\end{equation}
{At fixed $\omega>0$, the thermodynamic strong-coupling gap therefore grows with a leading power $|g|^{1/3}$, compared with $|g|^{1/2}$ in spin-wave theory.}

The mapping also exposes a continuous symmetry generated in the original axes by $\tfrac12\sum_j\epsilon_j\hat\sigma_j^z$. These are opposite rotations on the two sublattices, not the uniform axial rotations considered in Sec.~\ref{sec:symmetries}, so they do not forbid nearest-neighbor pair creation. Under the same transformation, the uniform $\hat S_y$ and $\hat S_z$ become staggered sums, whereas $\hat S_x$ is unchanged. The mapping fixes the spectrum exactly but does not identify the original collective squeezing with uniform-spin squeezing in the transformed model.

{\subsection{Odd numbers of spins}
\label{app:odd_xxz}

For odd periodic chains with $N>2$, the nearest-neighbor Hamiltonian and its normalization remain those of Eq.~\eqref{eq:nn_xxz}. The weak-coupling relations in Eq.~\eqref{eq:xxz_weak_coupling} apply to both even and odd $N$. Exact calculations for $N=7,9,11$ likewise find gap-opening squeezing at negative coupling. Their strong-coupling behavior requires a separate treatment. The alternating rotation used in Eq.~\eqref{eq:xxz_staggered_mapping} leaves the closing bond of an odd ring unrotated, so the Heisenberg mapping does not apply in its stated form.

To obtain the fixed-size strong-coupling limit, we treat the field as a perturbation to the interaction. For each of these odd rings, diagonalizing the interaction in sectors of fixed $\hat S_x$ gives an isolated ground doublet with eigenvalues $\pm1/2$. Denote these states by $|\Phi_\pm\rangle$. They need not belong to the maximal-spin sector. The field $2\omega\hat S_z$ has zero diagonal matrix elements in this doublet and a nonzero off-diagonal matrix element. It therefore selects an equal-weight superposition as its lower-energy state. As $g/\omega\to-\infty$ at fixed $N$ and $\omega>0$,
\begin{align}
\langle\hat S_z\rangle
&\longrightarrow-|\langle\Phi_+|\hat S_z|\Phi_-\rangle|,\nonumber\\
\langle(\Delta\hat S_x)^2\rangle
&\longrightarrow\frac14,\nonumber\\
\Delta_{\mathrm{mb}}
&\longrightarrow4\omega|\langle\Phi_+|\hat S_z|\Phi_-\rangle|.
\label{eq:odd_xxz_doublet}
\end{align}
Increasing $|g|$ shifts both doublet states by the same interaction energy, while the other levels move away from them. The lowest excitation consequently remains the field-induced splitting within the doublet, and the many-body gap saturates for these fixed odd sizes.

The selected state has its mean spin along $z$. For odd $N$, every transverse spin component has half-integer eigenvalues and zero mean, so its variance is at least $1/4$. The $\hat S_x$ variance in Eq.~\eqref{eq:odd_xxz_doublet} attains this minimum. The limiting Wineland parameter is therefore
\begin{equation}
\xiR^2\longrightarrow
\frac{N}{4|\langle\Phi_+|\hat S_z|\Phi_-\rangle|^2}.
\label{eq:odd_xxz_squeezing}
\end{equation}

This is the same mechanism as in the odd-$N$ Ising LMG limit (see Appendix~\ref{app:finite_spin} for its derivation). There, $|\Phi_\pm\rangle$ are the symmetric Dicke states with $S=N/2$ and $m_x=\pm1/2$, so $|\langle\Phi_+|\hat S_z|\Phi_-\rangle|=(N+1)/4$. Equations~\eqref{eq:odd_xxz_doublet} and~\eqref{eq:odd_xxz_squeezing} then reproduce the limits $\Delta_{\mathrm{col}}\to\omega(N+1)$ and $\xiR^2\to4N/(N+1)^2$ of the LMG model restricted to $S=N/2$. The all-to-all XXZ interaction has precisely this doublet. For nearest-neighbor interactions, spatial correlations change the matrix element, even though the squeezed variance still tends to $1/4$. Thus the common doublet mechanism does not by itself imply the LMG squeezing coefficient or its $1/N$ scaling.

For the nearest-neighbor ring with $N=9$, we obtain $|\langle\Phi_+|\hat S_z|\Phi_-\rangle|\simeq1.89934$, compared with $2.5$ in the symmetric Dicke doublet. Consequently,
\begin{equation}
\Delta_{\mathrm{mb}}/\omega\longrightarrow7.59735,
\qquad
\xiR^2\longrightarrow0.62370.
\label{eq:odd_xxz_limits}
\end{equation}
Full-spectrum diagonalization approaches these limits. As in the odd LMG model, stronger squeezing occurs before the doublet limit is reached. For example, $g/\omega=-10$ gives $\xiR^2\simeq0.41357$ and $\Delta_{\mathrm{mb}}/\omega\simeq7.30540$. Here the field also admixes higher interaction states, and the increased polarization more than compensates for the increased $\hat S_x$ variance. These results take the strong-coupling limit at fixed $N$, whereas Eq.~\eqref{eq:xxz_exact_asymptotic_gap} takes the thermodynamic limit first.}

\section{Nearest-neighbor Ising details}
\label{app:finite_range_benchmarks}

The nearest-neighbor limit of Eq.~\eqref{eq:powerlaw_ising} makes the difference between the collective response and many-body gap particularly clear. For a periodic chain with even $N>2$, the Kac factor is $\mathcal N_N=2$ and the Hamiltonian becomes
\begin{equation}
\hat H_{\mathrm I}^{\mathrm{nn}}=\omega\sum_i\hat\sigma_i^z
-\frac g2\sum_i\hat\sigma_i^x\hat\sigma_{i+1}^x.
\label{eq:nn_ising}
\end{equation}
The chain has $N$ nearest-neighbor bonds. Projecting their sum onto the maximal-spin sector gives
\begin{equation}
\hat P_{\mathrm{sym}}\hat H_{\mathrm I}^{\mathrm{nn}}\hat P_{\mathrm{sym}}
=\left(2\omega\hat S_z-\frac{2g}{N-1}\hat S_x^2\right)\hat P_{\mathrm{sym}}
+\mathrm{const.}
\label{eq:nn_ising_projection}
\end{equation}
Equation~\eqref{eq:nn_ising_projection} is exact within the projected sector {(see Appendix~\ref{app:xxz_checks} for the symmetric-pair identity)}, but does not give the complete low-energy spectrum. It retains the uniform fluctuation and omits spatially varying modes. Positive $g$ then lowers the energy cost of $\hat S_x$ fluctuations and squeezes $\hat S_y$. Negative $g$ raises that cost, squeezes $\hat S_x$, and increases the uniform excitation energy. The projection explains the rotation of the squeezed axis but cannot determine the many-body gap.

Comparing Eq.~\eqref{eq:nn_ising_projection} with Eq.~\eqref{eq:lmg} gives $g_{\mathrm{LMG}}=gN/(N-1)$. At quadratic order, the projected stiffnesses are therefore $K_X=2[\omega-gN/(N-1)]$ and $K_P=2\omega$.

The spatially varying modes appear in a local Holstein--Primakoff expansion,
\begin{align}
\hat H_{\mathrm I}^{(2)}={}&\sum_k\left(2\omega-g\cos k\right)
\hat a_k^\dagger\hat a_k\nonumber\\
&-\frac g2\sum_k\cos k
\left(\hat a_k\hat a_{-k}+\hat a_k^\dagger\hat a_{-k}^\dagger\right),
\label{eq:nn_ising_bdg}\\
\varepsilon_{\mathrm I}^{\mathrm{sw}}(k)={}&2\sqrt{\omega\left(\omega-g\cos k\right)}.
\nonumber
\end{align}

The $k=0$ mode is the uniform fluctuation probed by $\hat S_x$ and $\hat S_y$. It softens for positive $g$ and hardens for negative $g$. The $k=\pi$ mode instead alternates between sublattices and softens for $g<0$. It is absent from Eq.~\eqref{eq:nn_ising_projection}, which retains only the uniform mode. The sublattice rotation
\begin{equation}
\hat U_{\mathrm{st}}=\prod_{j\,\mathrm{even}}\hat\sigma_j^z
\end{equation}
maps $g$ onto $-g$ while transforming a uniform transverse spin into a staggered one. The spectra for the two signs are identical, although the soft mode moves from momentum $0$ to $\pi$. The quadratic expansion captures this exchange but not the exact transition point of the one-dimensional spin-$1/2$ chain. The Jordan--Wigner dispersion~\cite{LiebSchultzMattis1961,Pfeuty1970,BarouchMcCoy1971} is
\begin{equation}
\varepsilon_{\mathrm I}(k)=2\sqrt{\omega^2+\frac{g^2}{4}-\omega g\cos k}.
\label{eq:nn_ising_exact}
\end{equation}
It closes at $g=2\omega$ and $k=0$, or at $g=-2\omega$ and $k=\pi$. In the thermodynamic paramagnetic phase, the many-body gap is $\Delta_{\mathrm{mb}}=2\omega-|g|$ and decreases for either sign of the coupling. Thus, the nearest-neighbor Ising chain has no anti-critical branch when the gap is the lowest physical excitation. Negative coupling hardens the uniform mode, while a staggered excitation moves down and eventually closes the many-body gap. This distinction also explains the saturation of uniform squeezing (see Appendix~\ref{app:ising_saturation} for the derivation).

\section{Saturation of uniform squeezing in the nearest-neighbor Ising chain}
\label{app:ising_saturation}

The momentum distinction has a direct consequence for the Wineland parameter because the total spin probes only zero momentum. {For the translation-invariant Ising ground state considered here, with nonzero mean spin along $z$, let $m_z=\langle\hat\sigma_i^z\rangle=2\langle\hat S_z\rangle/N$.} We use $\langle\hat A\hat B\rangle_c=\langle\hat A\hat B\rangle-\langle\hat A\rangle\langle\hat B\rangle$ for connected correlations and define the transverse structure factor
\begin{equation}
\mathcal S_{\mu\mu}(q)=\sum_r e^{-\ii q r}
\langle\hat\sigma_0^\mu\hat\sigma_r^\mu\rangle_c.
\end{equation}
Here $q$ is the wave number. Choosing the transverse principal axis $\mu=x$ or $y$ with the smaller variance, translation invariance gives
\begin{equation}
\xiR^2=\frac{\mathcal S_{\mu\mu}(0)}{m_z^2}.
\label{eq:nn_structure_squeezing}
\end{equation}
In the paramagnetic phase, away from $|g|=2\omega$, the connected transverse correlations decay exponentially~\cite{Pfeuty1970,BarouchMcCoy1971}. {Their zero-momentum sum $\mathcal S_{\mu\mu}(0)$ and the polarization $m_z$ therefore approach finite values as $N$ grows, giving $\xiR^2\sim N^0$.} This conclusion also holds for the minimum-variance uniform component at criticality. At $g=2\omega$, the $y$ correlations decay as $r^{-9/4}$ and remain summable. At $g=-2\omega$, the slowly decaying $x$ correlations are staggered as $(-1)^r r^{-1/4}$. {The resulting singularity is in $\mathcal S_{xx}(q)$ at $q=\pi$, not at the uniform momentum $q=0$ entering Eq.~\eqref{eq:nn_structure_squeezing}}~\cite{Pfeuty1970,BarouchMcCoy1971}.

In a parity-preserving ordered state, the order-parameter correlations retain a long-range part, so an exponential-decay argument does not apply to every transverse component. For $g>2\omega$, the minimum-variance uniform component is $y$, whose correlations remain short ranged~\cite{BarouchMcCoy1971}. For $g<-2\omega$, the long-range $x$ correlations are staggered. Their ordered contribution lies at momentum $\pi$ and cancels from the uniform sum for an even periodic chain. After separating this contribution, the remaining correlations are summable. Thus, at fixed finite $g/\omega$ and nonzero $m_z$, the minimum collective variance grows proportionally to $N$ and the Wineland parameter approaches an $N$-independent value.

{This saturation is specific to uniform squeezing in the Ising chain, not a general restriction on short-range models~\cite{Comparin2022,CalecaLuttinger2025}.}

\section{Numerical definitions}
\label{app:numerics}

We give the numerical grids and diagonalization procedures for the power-law and finite-temperature results.

{Figures~\ref{fig:XXZ} and~\ref{fig:finite_range_ising} each use $61$ coupling values, comprising the $51$ uniformly spaced values specified below and ten additional values near zero. Figure~\ref{fig:xxz_thermal} uses $132$ coupling values, comprising $121$ uniformly spaced values and eleven additional values near zero. The additional nonzero-coupling points use complete diagonalization of both parity blocks. The $g=0$ values are included analytically, with $\xiR^2=1$ and $\Delta_{\mathrm{mb}}=\Delta_{\mathrm{col}}=2\omega$ at zero temperature, and $\xiR^2=\coth^2(\omega/T)$ at positive temperature.}

\subsection{XXZ interaction and finite-temperature scans}

For Fig.~\ref{fig:XXZ}, we use the periodic distance and Kac normalization of Eqs.~\eqref{eq:powerlaw} and~\eqref{eq:kac}. The $N=10$ base grid contains $51$ values of $g/\omega\in[-10,2]$ and $51$ values of $\alpha\in[0,4]$. We obtain the ground state and many-body gap from the two parity sectors and retain $256$ states from each sector to evaluate Eq.~\eqref{eq:collective_gap_definition}. These states resolve at least $99.60\%$ of the exact total transverse weight $\sum_{E>E_0}W^\perp(E)$, fixed by the sum rule in Eq.~\eqref{eq:collective_sum_rule}. We find $\Delta_{\mathrm{mb}}=\Delta_{\mathrm{col}}$ at every grid point to machine precision. The first excited level contributes $W^\perp(E_1)$, at least $86.18\%$ of that same exact total. The first percentage tests numerical completeness, whereas the second measures the fraction of the response at $E_1$.

For Fig.~\ref{fig:xxz_thermal}, we use Eq.~\eqref{eq:nn_xxz} with $N=10$ and $k_{\mathrm B}=1$. Thermal moments are evaluated in the exact Gibbs state
\begin{equation}
\hat\rho_T
=\frac{e^{-\hat H_{\mathrm{XXZ}}^{\mathrm{nn}}/T}}
{\operatorname{Tr}e^{-\hat H_{\mathrm{XXZ}}^{\mathrm{nn}}/T}}.
\label{eq:thermal_state}
\end{equation}
The $T=0$ endpoint is evaluated from the ground state.

The base grid contains $121$ values of $g/\omega\in[-6,1]$ and $121$ values of $T/\omega\in[0,4.5]$. At each coupling, we diagonalize both parity blocks completely, combine their spectra, and evaluate the Gibbs trace without truncation. Parity forces the transverse mean spin to vanish. The Hamiltonian is real in the {product basis of the local $\hat\sigma_i^z$ operators}, so the symmetrized $\hat S_x$--$\hat S_y$ covariance vanishes and these operators are the transverse principal axes. We check the implementation against the analytic noninteracting result $\xiR^2=\coth^2(\omega/T)$ and an independent full-density-matrix covariance calculation.


\subsection{Size and coupling-budget checks of the thermal comparison}

{For the two fixed grid couplings $g/\omega=8/15$ and $-25/24$, Table~\ref{tab:thermal_size_check} compares complete-spectrum results for $N=6,8,10,12$. We locate the squeezing-loss temperature $T_*$ defined in Sec.~\ref{sec:thermal_robustness} by bisection using the exact Gibbs moments. The labels $+$ and $-$ distinguish the positive- and negative-coupling states. Both $\xiR^2(T=0)$ and $T_*$ show little size dependence at these parameters.}

\begin{table}[tbp]
\caption{Full-spectrum checks for the nearest-neighbor XXZ chain. The labels $+$ and $-$ refer to $g/\omega=8/15$ and $-25/24$, respectively. The two columns headed $\xiR^2$ give zero-temperature values.}
\label{tab:thermal_size_check}

\begin{ruledtabular}
\begin{tabular}{ccccc}
$N$ & $\xiR^2(+)$ & $\xiR^2(-)$ & $T_*^{(+)}/\omega$ & $T_*^{(-)}/\omega$ \\
6  & 0.578103 & 0.578615 & 0.493005 & 1.584655 \\
8  & 0.577142 & 0.578807 & 0.493528 & 1.584773 \\
10 & 0.577048 & 0.578824 & 0.493571 & 1.584777 \\
12 & 0.577039 & 0.578825 & 0.493574 & 1.584777
\end{tabular}
\end{ruledtabular}

\end{table}

To specify a second resource comparison, count the absolute coefficients of the Pauli terms in Eq.~\eqref{eq:nn_xxz} and divide their sum by $N$. There are $N$ field terms of magnitude $\omega$ and $3N$ bond terms of magnitude $|g|/2$, giving $\Lambda=\omega+3|g|/2$. Comparing at a common value of $\Lambda$ rescales the field as well as the interaction, while leaving $g/\omega$ and ground-state squeezing unchanged. For the two $N=10$ states, $\Lambda/\omega=1.8$ and $2.5625$, respectively. Their thresholds are then
\begin{equation}
T_*^{(+)}/\Lambda\simeq0.274206,
\qquad T_*^{(-)}/\Lambda\simeq0.618450.
\end{equation}
{The ratio of the rescaled thresholds $T_*^{(-)}/\Lambda$ and $T_*^{(+)}/\Lambda$ is about $2.26$, compared with $T_*^{(-)}/T_*^{(+)}\simeq3.21$ at fixed $\omega$.}

\subsection{Parity-resolved finite-range spectrum}

For Fig.~\ref{fig:finite_range_ising}, we diagonalize the $N=10$ Hilbert space in the exact even and odd eigenspaces of $\hat\Pi_z$. This prevents arbitrary mixtures of nearly degenerate parity states in the ordered regime. The base grid uses $51$ values of $g/\omega\in[-10,2]$ and $51$ values of $\alpha\in[0,4]$. At each point, we use the global ground state to evaluate the full transverse covariance matrix and the many-body gap. We retain $256$ states from each parity sector for Eq.~\eqref{eq:collective_gap_definition}. {They resolve at least $97.14\%$ of the exact total transverse weight $\sum_{E>E_0}W^\perp(E)$ in Eq.~\eqref{eq:collective_sum_rule} over the full grid.}

\bibliography{references_v14}

\end{document}